\documentclass[journal]{IEEEtran}

\usepackage{amssymb}
\usepackage{latexsym}

\usepackage{url}
\usepackage{xcolor}
\usepackage{booktabs}
\usepackage{color,soul}

\usepackage{amsmath}
\usepackage{mathtools}
\usepackage{enumitem}

\usepackage{dblfloatfix}
\usepackage{tikz}       
\usepackage{nicematrix} 
\usepackage{framed,multirow}

\usepackage{booktabs,arydshln}
\makeatletter
\def\adl@drawiv#1#2#3{%
        \hskip.5\tabcolsep
        \xleaders#3{#2.5\@tempdimb #1{1}#2.5\@tempdimb}%
    #2\z@ plus1fil minus1fil\relax
        \hskip.5\tabcolsep}
\newcommand{\cdashlinelr}[1]{%
  \noalign{\vskip\aboverulesep
           \global\let\@dashdrawstore\adl@draw
           \global\let\adl@draw\adl@drawiv}
  \cdashline{#1}
  \noalign{\global\let\adl@draw\@dashdrawstore
           \vskip\belowrulesep}}
\makeatother

\usepackage{lipsum}
\usepackage[switch]{lineno}
\usepackage{tcolorbox}
\usepackage[colorlinks=true,linkcolor=red]{hyperref}

\usepackage[labelsep=period,labelfont=bf]{caption}
\usepackage[nameinlink]{cleveref}  
\crefname{figure}{Fig.}{\textbf{Figure.}}
\crefname{equation}{Eq.}{\textbf{Eq.}}
\crefname{table}{Table}{\textbf{Table.}}
\crefname{section}{Section}{\textbf{Section}}

\newcommand{\lvl}{~~~~}

\definecolor{newcolor}{rgb}{.8,.349,.1}

\hypersetup{
	colorlinks,
	linkcolor=[rgb]{1.0, 0.0, 0.0},
	citecolor=[rgb]{1.0, 0.2, 0.2},
	urlcolor =[rgb]{0.0, 0.0, 0.8}
}

\begin{document}

\title{Handcrafted Histological Transformer (H2T): Unsupervised Representation of Whole Slide Images}

\author{~Quoc~Dang~Vu$^1$, ~Kashif~Rajpoot$^2$, ~Shan~E~Ahmed~Raza$^1$
and~Nasir~Rajpoot$^{1,*}$\\ $^1$ \{quoc-dang.vu, shan.raza, n.m.rajpoot\}@warwick.ac.uk \\ $^2$ k.m.rajpoot@bham.ac.uk \\ $^*$ Corresponding author
\thanks{Q.D.Vu, S.E.A.Raza and N.Rajpoot are from the Tissue Image Analytics Centre, Department of Computer Science, University of Warwick, UK}
\thanks{K.Rajpoot is from the School of Computer Science, University of Birmingham, UK}
\thanks{N.Rajpoot is also affiliated with The Alan Turing Institute, London, UK and the Department of Pathology, University Hospitals Coventry \& Warwickshire, UK}
}

\maketitle

\begin{abstract}
Diagnostic, prognostic and therapeutic decision-making of cancer in pathology clinics can now be carried out based on analysis of multi-gigapixel tissue images, also known as whole-slide images (WSIs). Recently, deep convolutional neural networks (CNNs) have been proposed to derive unsupervised WSI representations; these are attractive as they rely less on expert annotation which is cumbersome. However, a major trade-off is that higher predictive power generally comes at the cost of interpretability, posing a challenge to their clinical use where transparency in decision-making is generally expected. To address this challenge, we present a handcrafted framework based on deep CNN for constructing holistic WSI-level representations. Building on recent findings about the internal working of the Transformer in the domain of natural language processing, we break down its processes and \textit{handcraft} them into a more transparent framework that we term as the Handcrafted Histological Transformer or H2T. Based on our experiments involving various datasets consisting of a total of 10,042 WSIs, the results demonstrate that H2T based holistic WSI-level representations offer competitive performance compared to recent state-of-the-art methods and can be readily utilized for various downstream analysis tasks. Finally, our results demonstrate that the H2T framework can be up to 14 times faster than the Transformer models.
\end{abstract}


\IEEEpeerreviewmaketitle

\section{Introduction}

Visual assessment of tissue specimens under the microscope remains the {\em gold standard} for diagnosis of cancer and used for the purposes of prognostication and therapeutic planning \cite{Gurcan2009GoldStandard, Abels2019DPathWhitePaper}. With the advancement in digitization, current pathology workflows increasingly use multi-gigapixel tissue images, now commonly known as whole slide images (WSIs), in a wide range of settings. These WSIs also enable pathologists to view the tissue samples remotely. Computational analysis of WSIs offers the promise for the detection of known diseases and, perhaps, the discovery of new disease subtypes.

In recent years, several machine learning approaches have been proposed for identifying nuclei, glandular structures or tumor-rich regions in histology images \cite{Simon2020HoVeNet,Simon2020Steerable,Kumar2021Monusac,kather2019predicting}. There are currently two major approaches for WSI-level analysis. The first one is to construct features based on classification, detection or segmentation of the tissue components. These features are typically designed based on our knowledge from biological findings \cite{Gentles2015ImmuneCells}, such as the co-localization of lymphocytes surrounding cancerous epithelium \cite{Shaban2019TILScore} or the deformation of glands in colon samples \cite{Awan2017GlandularFeatures}. Despite their effectiveness in prognosis and providing interpretability, there are several drawbacks during the construction of such pipelines. First and foremost, they rely mostly on annotated samples which are often intensive in terms of expert pathologists' time and effort \cite{amgad2019CrowdSourceAnnotation, kromp2020annotated}. In addition, the pathologists are well-known to have high discordance on how a tissue sample or its constituents are labelled \cite{Ayesha2021PathologistConcordance, wahab2021SemanticAnnotation}.

{
In contrast, recent approaches have focused more on improving the discriminative power of the features \cite{lu2020CLAM, Li2021DSMIL} rather than on the mechanism to derive a generic representation at the WSI-level. Although these approaches have achieved promising results, as with most deep learning based methods, they lack transparency and interpretability for their predictions. To mitigate this, recent techniques have utilized the attention mechanism to output a heatmap to indicate which instances the models rely on for making predictions \cite{Welling2018AMIL, lu2020CLAM}.
}

In this paper, we propose a novel way to obtain unsupervised holistic WSI-level representations based on a set of data-driven histological patterns, which we term as the \textit{histological prototypical patterns}. The proposed representations, which we term as the handcrafted histological transformer (H2T) representations, are inspired by the attention mechanism of the well-known Transformers in natural language processing (NLP) \cite{Vaswani2017Transformer} and attempt to model the attention mechanism in a handcrafted manner. We show that the proposed H2T representations are discriminative and can be readily utilized for various downstream analysis tasks with significantly reduced amount of effort. These representations are mined from the pixel data in WSIs and {\em handcrafted} from deep feature based representations or co-localization of histological pattern maps that commonly appear throughout the WSIs while indirectly incorporating the attention mechanism. Similar to the features constructed from tissue components which are considered to be highly interpretable, these patterns also facilitate tractability and interpretability of our derived WSI representations compared to other methods. We demonstrate that such interpretations can be achieved either through visual assessment or by retrieving closest image patches to the prototypical patterns. We evaluate the capacity of the derived prototypical patterns and the resulting H2T representations using two large publicly available WSI datasets: The Cancer Genome Atlas (TCGA) and Clinical Proteomic Tumour Analysis Consortium (CPTAC).

The main contributions of this work are as follows:

\begin{itemize}

\item We present a novel paradigm, termed as the handcrafted histological transformer (H2T), for deriving holistic WSI-level representations;
\item We show how the proposed H2T representations can be constructed from histological prototypical patterns that are mined from WSIs in an unsupervised manner; in addition, we show how the prototypical patterns can be interpreted biologically and later utilized for discovery purposes;
\item We provide a baseline Transformer model for WSI-level analysis, the first of its kind to the best of our knowledge;
\item We show that the H2T representations are as predictive as the recent state-of-the-art methods (including the aforementioned Transformer model) while being computationally much cheaper, based on results from experiments on {6 datasets consisting of a total of 10,042 WSIs}.
{\item We provide the code and the intermediate data at \url{https://github.com/vqdang/H2T} to facilitate future investigation efforts.}
\end{itemize}

\section{Related work}

\subsection{Handcrafting representations for cytology and histology images}

From the clinical and biological findings thus far, the morphology and distribution of tissue components such as gland or nuclei are recognised as strong indicators for cancer patient survival  \cite{elston1991NuclearPleomorphism, Goldstraw2016LungStaging, Gentles2015ImmuneCells}. In lung tissue for instance, micropapillary and solid pattern are related to cancer with high degree of aggressiveness \cite{solis2012LungHistologic, cao2016LungMicropapillaryClinical}.

Early automated systems attempted to utilize the above information to differentiate tumor from normal tissue images. \cite{Hamilton1994} proposed a Bayesian network using the amount of nuclei, nuclear size, mucinous area, etc. quantified by cytologists as features for prediction. \cite{Hamilton2000} developed a primitive automated nuclei detection method and then employed Delaunay triangulation to characterize the spatial distribution of detected nuclei. In their work, the resulting statistics of edges, vertices and triangles were features for predicting cervical intraepithelial neoplasia. 

With the increase in compute power, automated methods which were previously restricted to just small images were then extended to tissue microarray (TMAs) and WSIs. \cite{Tabesh2007, JTKwak2017TMA-Handcrafted-Features} utilized morphological and textural features obtained from image patches for stratifying tumor grades of TMA cores. On the other hand, \cite{sertel2009computer} utilized a multi-resolution approach and extracted textural features at all resolution levels and then utilized them for making and refining prediction on WSIs in a coarse-to-fine manner.

Nonetheless, even with deep learning, processing large images by and large is still achieved by breaking them down into smaller parts (i.e., patches). In \cite{Hou2016}, the authors first employed convolutional neural networks (CNNs) to classify image patches. Later, they either utilized the resulting histogram or fused features from patches to make predictions for an entire WSI. Around the same time, new clinical findings indicated that a large number of lymphocytes infiltrating deep within the tumor sites carry prognostic significance \cite{shield2017TILClinical}. Shortly after, \cite{Shaban2019TILScore} proposed an automated method to obtain a single score to measure the amount of tumor-infiltrating lymphocytes (TILs) that is predictive of patient survival. This involves robustly identifying the tissue types of all image patches within a large cohort of WSIs. This form of analysis is taken further in later works. \cite{Diao2021HandcraftWSI-Mutation} extracted multiple statistics of tissue components and demonstrated that these features still correlate well with recently established tumor microenvironment markers as well as molecular signatures.

\subsection{Learning WSI representations}

While handcrafted WSI features like the above remain a potent way to predict disease status \cite{amgad2019CrowdSourceAnnotation, kromp2020annotated}, the methods for obtaining those representation are often laborious and time-consuming. To lessen the burden of annotation, the medical image processing analysis community turned to multiple instance learning (MIL) to predict the label of a bag of instances without needing to identify the labels of the constituent parts (or instances) \cite{dietterich1997MIL, andrews2002MIL-SVM}. {In computational pathology, each instance is a feature vector of an image patch. These vectors are assumed to be highly compact while still being discriminative enough for major tissue patterns. Due to this assumption and heavy reliance on deep feature representation of a patch, a majority of the techniques have focused more on improving the discriminative power of the features \cite{lu2020CLAM, Li2021DSMIL, kalra2021PAF, Abbet2020DivideandRuleSL}. In particular, several methods applied weakly supervised learning while treating the WSIs as bags of instances (image patches) with respect to a specific task, such as classification of WSIs \cite{Fuchs2019Nature, mohsin2021MSILancet}}.

While such systems could allow us to do away with a large amount of human a \textit{priori} knowledge, finding and attributing which instances are important to the prediction remains difficult \cite{kandemir2015MIL-Benchmark}. Without being able to localize down to the instance-level, interpretation of the model could not be made for clinical settings. In order to resolve this, recent works like \cite{Fuchs2019Nature} utilized visualization techniques to increase the interpretability of their results. Specifically, while they used a recurrent neural network (RNN) for prediction, they applied t-SNE \cite{van2008TSNE}, a manifold mapping, on their input instances to extract their placements within the model decision space.

Recently, neural networks with attention mechanism came forth as a powerful tool for the medical image analysis community. In particular, they can not only learn a discriminative bag-level representation but also provide interpretable and relatable instance-level attributions \cite{Welling2018AMIL}. In order to apply this method to WSI-level analysis, WSIs are commonly split into patches where each is passed through a pretrained CNN for feature extraction. The resulting set of feature vectors is then input to a proposed neural network for training and making predictions \cite{lu2020CLAM, Li2021DSMIL, kalra2021PAF, Noriaki2020MS-ABMIL} while each image patch is considered as an instance in the MIL setting. Because the patch-level representation is the most critical building-block for these methods, many works focus on designing a scheme to enrich this representation. MS-DA-MIL \cite{Noriaki2020MS-ABMIL} employed adversarial training to enrich the instance-level features before inputting them into a multi-scale attention model. More recently, CLAM \cite{lu2020CLAM} utilized a loss to pull the representation of patches within a specific label closer together. DSMIL-LC \cite{Li2021DSMIL} proposed using SimCLR \cite{chen2020SimCLR} (self-supervised contrastive learning) to derive patch-level representations and a scheme to combine instance-level prediction with bag-level prediction. FocAtt-MIL \cite{kalra2021PAF} refined the patch-level features by training the feature extractor with hierarchy of instance labels.

Concurrent to the above developments, natural language processing (NLP) recently experienced tremendous breakthrough via the adoption of Transformer models \cite{Vaswani2017Transformer} which were pretrained on large NLP datasets such as GPT-3 \cite{Tom2020GPT3}. Although Transformer is also an attention-based neural network at its core, we consider it as a more generalized form of attention mechanism compared to the methods mentioned previously. Since its inception, Transformers have been adopted by other fields with great success. For instance, using a Transformer-based architecture, AlphaFold achieved a significant improvement compared to all other methods on a 50 years old grand challenge \cite{jumper2021AlphaFold} in predicting the protein folding structure. In natural image analysis, Vision Transformers have achieved comparable performance against ResNet \cite{Dosovitskiy2020VisionTransformer}. Furthermore, in some cases, Transformer-based networks have shown to be more robust than CNNs \cite{hendrycks2021natural}. In computational pathology, \cite{myronenko2021PandaTransformer} have recently adopted Transformer for predicting Gleason grades of WSIs.

{
Recent theoretical analysis and empirical evidence on Transformer families (i.e those utilizing multi-head self-attention mechanism) have demonstrated that they have strong capability for retrieving information. Specifically, the theoretical analysis by \cite{Ramsauer2021Hopfield} showed that MHA can directly use raw images for querying and/or storing other raw images (i.e. acting like code books in term of dictionary learning) without any training. On the other hand, \cite{el2021training} shows that Transformer can be effectively turned into a strong image retrieval system by simply using a different loss function. Lastly, \cite{caron2021emerging} has shown that a frozen Vision Transformer can be used ``out of the box” for image retrieval. With this evidence, we consider Transformer as one of the most powerful context-based image retrieval (CBIR) techniques in the current times for representing WSIs. 
}

{
\subsection{Unsupervised learning}
Traditionally, unsupervised learning methods are defined as techniques that do not rely on labels to obtain the data underlying representation. Prime examples include clustering methods and other traditional dimensionality reduction techniques like Principal Component Analysis, k-means clustering, UMAP \cite{mcinnes2018umap} or t-SNE \cite{van2008TSNE}. However, in the current literature, by relaxing the definition of ``no label'' to ``\textit{\textbf{no human} supervisory signals}'' \cite{doersch2017multi}, unsupervised learning can be framed as self-supervised learning. Thus, self-supervised learning is a subset of unsupervised learning where the guidance signals for the training process can be obtained by directly interacting with the data itself. Typical self-supervised learning supervisory signals include filling in image holes, solving jigsaw puzzles made from image patches, predicting movement in videos or more \cite{kim2019self,chen2020SimCLR}. Empirically, relaxing the conditions about the label origins have led to the discoveries of much more general and robust representations \cite{caron2020SWAV, chen2020SimCLR}.
}

{
In relation to H2T, while our proposed framework's performance relies on the deep features, which can be obtained either by self-supervised learning or supervised learning, its internal mechanism consists of only clustering methods and operations that do not rely on any human labels. Thus, we consider that our proposed framework fits well in the traditional category of unsupervised learning as described previously.
}

\subsection{Representation of image patches}

Several applications, not limited to medical image processing, have utilized CNNs pretrained on ImageNet in a supervised manner for various different tasks. However, recent advancements in computer vision have again emphasized the importance of obtaining a strong representation. In particular, \cite{hendrycks2019using, djolonga2021robustness, navid2021self} have demonstrated that representations derived from self-supervised learning are more robust than those obtained solely from supervised learning. Many recently proposed techniques like SimCLR \cite{chen2020SimCLR} are also rapidly closing the gap between self-supervised learning and supervised learning. In particular, SWAV \cite{caron2020SWAV} surpassed the performance of ResNets which were trained on ImageNet in a supervised manner. In computational pathology, \cite{ciga2020self} recently assessed SimCLR on a large cohort and showed that self-supervised training small CNNs on histopathology images is more beneficial for downstream tasks compared to those pretrained on ImageNet with supervised learning.

\begin{figure*}[!t]
\centering
\includegraphics[width=0.9\textwidth]{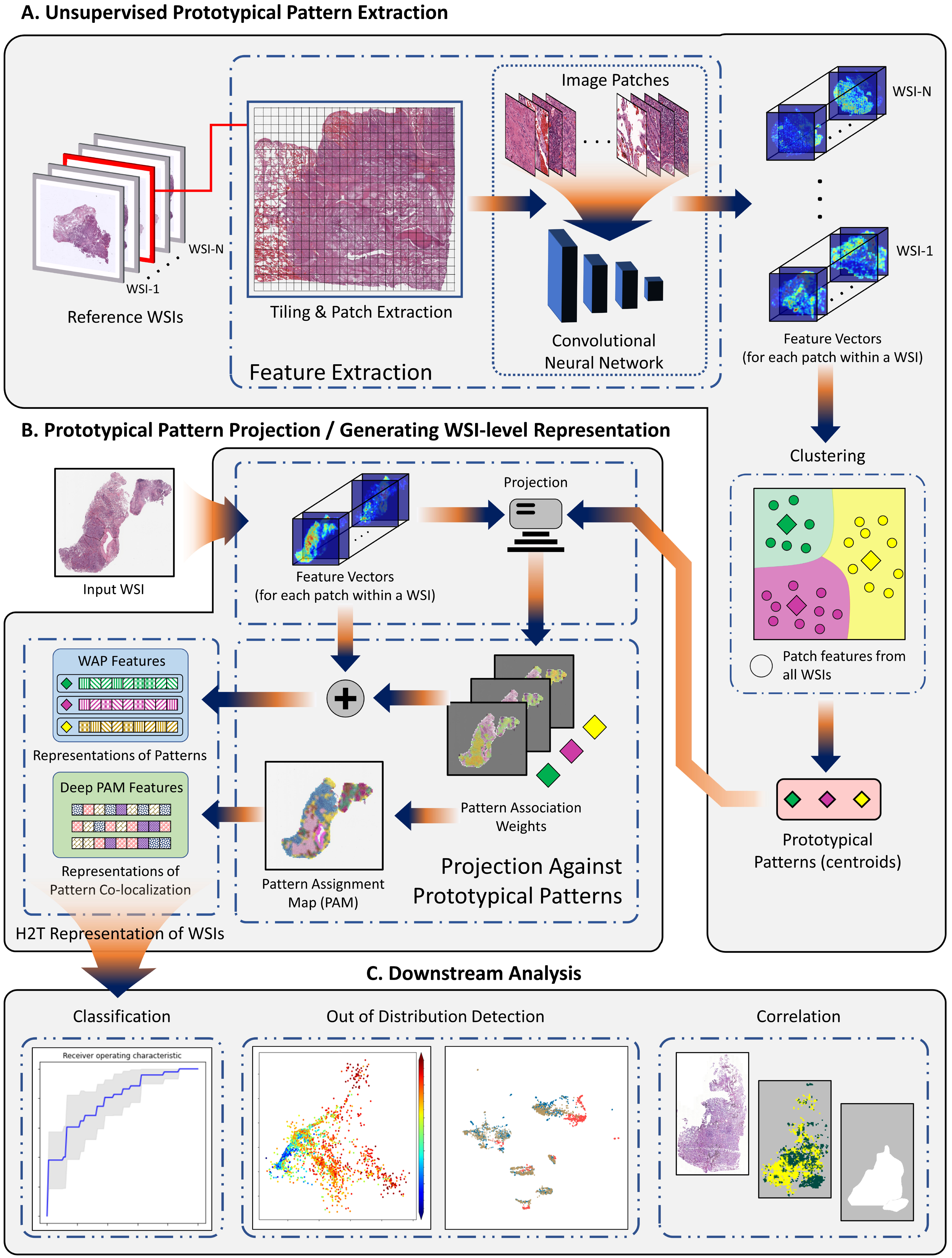}
\caption{Our proposed Handcrafted Histological Transformer (H2T) framework. The framework revolves around the extraction and the utilization of the prototypical patterns. By projecting a new WSI against this set of prototypical patterns, we derive highly discriminative WSI representations that are readily usable for other downstream tasks. In the feature vector images (the stacks of blue images), each pixel corresponds to a patch in the WSI and the depth corresponds to the features from a CNN. Throughout the framework, these features are extracted using the same pretrained CNN. In our case, this can either be from ResNet50 pretrained on ImageNet or ResNet50 pretrained by SWAV self-supervised learning method \cite{caron2020SWAV}. The Pattern Association Weights are described in \cref{eq:Hi_formula}.}
\label{f:pipeline}
\end{figure*}

\section{Methodology}

Recent automated methods not only require much less human annotation but also can be more predictive compared to their more traditional counterparts. However, as a trade-off for their improvement in predictive power, their internal processes are difficult to interpret to human operators. Moreover, their computational cost can sometimes be prohibitively expensive. Herein, we propose a method that is more interpretable and computationally cheaper without compromising on predictive power.

\subsection{Handcrafted histological transformer (H2T)}

Inspired by the Transformer and the recent works that unravel its mechanism, we breakdown Transformer operation and re-construct them in a handcrafted manner for histology-related tasks. We refer to the proposed framework as Handcrafted Histological Transformer (H2T). As shown in \cref{f:pipeline}, there are two stages of the H2T for representation learning:

\begin{enumerate}[label=\alph*.]
\item Construction of the prototypical patterns;
\item Projection against these patterns.
\end{enumerate}

In the first stage, we extract a set of prototypical patterns from a set of \textit{reference} WSIs (later referred to as the reference cohort). In order to obtain the most representative patterns, it is crucial to utilize a large enough and representative repository of WSIs from multiple sources. In the second stage, new WSIs are projected against the prototypical patterns and are summarized based on the relationship between their constituent instances and their assigned patterns. In the final stage, we utilize the resulting representation of WSIs for subsequent analysis.

We demonstrate that the resulting representations are highly discriminative and can be readily used with relative ease. Owing to the relatively low computational requirements, the predictive power of their representations and many unsupervised steps within, we show that the proposed framework can also be used for data discovery purposes such as out of distribution detection.

In the remainder of this section, we first describe the key mechanism behind the Transformer attention. We then describe in depth how H2T representations are formulated in a similar fashion but without employing an explicit attention module.

\subsection{Multi head self-attention}

The multi head (self) attention (MHA or MHSA) architecture and its powerful modeling capacity was popularized via the Transformer architecture \cite{Vaswani2017Transformer}. The core of the Transformer, or the MHA to be exact, is centered around the following formulation:
\begin{equation}
\label{eq:MHA}
\begin{split}
\widehat{Q} & = softmax \left( \frac{1}{\sqrt{d_k}} Q K^T \right) V \\
            & = softmax \left( \frac{1}{\sqrt{d_k}} Q W_Q W_K^T K^T \right) V W_V \\
\end{split}
\end{equation}
\noindent
Here, $\widehat{Q}$ is the attention output of a single head while $K$, $Q$ and $V$ are commonly referred to as the key, query and value inputs. We denote the associated dimensions of their features as $d_k$, $d_q$ and $d_v$. Additionally,  $W_K\in\mathbb{R}^{d_k \times d_e}$, $W_Q\in\mathbb{R}^{d_q \times d_e}$ and $W_V\in\mathbb{R}^{d_v \times d_e}$ are learnable weights for projecting each input feature into a common space with dimensionality $d_e$.

By constructing multiple such modules and selecting features within $\widehat{Q}$ from each head $h$ together, we obtain the renowned MHA architecture:
\begin{equation}
MultiHeadAttenion(Q, K, V) = Concat(\widehat{Q}_1, ..., 
\widehat{Q}_h) W_L
\end{equation}
\noindent
where $h$ is the total number of heads and the projection weight matrix $W_L\in\mathbb{R}^{h d_e\times d_v}$ is learnable.

According to \cite{Ramsauer2021Hopfield}, by using the same input $Y$ for $K$ and $V$ and by renaming the input $Q$ as $R$, \cref{eq:MHA} can take the form:

\begin{equation}
\label{eq:MHA_HopField}
\begin{split}
\widehat{Q} & = softmax(\frac{1}{\sqrt{d_k}} Q W_Q W_K^T K^T) V W_V \\
            & = softmax(\beta R W_Q W_K^T Y^T) Y W_V \\
\end{split}
\end{equation}
\noindent where $\beta$ is a scaling factor.

Under the above formulation, \cite{Ramsauer2021Hopfield} proved that MHA is closely related to Hopfield neural network. Additionally, they identified several interesting properties. First of all, the above equation is synonymous with finding the association between inputs $R$ and $Y$, using $R$ as reference. Secondly, the scaling factor $\beta=\sqrt{1 / d_k}$ is of particular importance as it controls the degree of memorization and association capacity of the architecture. Finally, if we take a step further and consider that $R$ is trainable, we effectively obtain an architecture that learns a set $P$ of prototypical patterns from the training set. Accordingly, this is synonymous to letting the network learn how all instances within the input are related to a prototypical pattern $p$. Once the set of patterns are identified, the network then performs weighted average pooling over input instances to derive a representation of the input sequence. Furthermore, by stacking multiple attention heads, it becomes possible for the network to derive multiple prototypical patterns.

\subsection{Positional encoding}

Another component that is often used together with MHA is the positional encoding. It is apparent from \cref{eq:MHA} that MHA is permutation invariant with respect to the ordering (or position) of input instances. As such, in cases where positions are of extreme importance, it is crucial for us to incorporate this information within the network design. At the moment, this is commonly achieved via sine-encoding (or Fourier-encoding) where they are either added or concatenated together with the instance features. In computer vision, using positional encoding makes a significant difference in performance for methods using MHA within their solution \cite{Carion2020DETR}.

With $d_{\psi}$ as the number of features (or embedding dimensions) within the vector representing the image patch (or instance), for 2D dimensions with $x$ and $y$ respectively as the instance positions along the $x$ and $y$ axes within the WSI, we use the following position encoding function (PE) to encode the position of each instance $\psi$ for a given embedding dimension $j$:
\begin{equation}
\label{eq:positional_encoding}
\begin{split}
PE(x, y, j) = Concat(PE_{sin}(x, 4j),PE_{cos}(x, 4j+1), \\
PE_{sin}(y, 4j+2),PE_{cos}(y, 4j+3))
\end{split}
\end{equation}
\begin{equation}
\begin{aligned}
PE_{sin}(x, 4j)   &= sin \left( \frac{x}{\epsilon^{4j/d_{\psi}}} \right) \\
PE_{cos}(x, 4j+1) &= cos \left( \frac{x}{\epsilon^{(4j+1)/d_{\psi}}} \right) \\
PE_{sin}(y, 4j+2) &= sin \left( \frac{y}{\epsilon^{(4j+2)/d_{\psi}}} \right) \\
PE_{cos}(y, 4j+3) &= cos \left( \frac{y}{\epsilon^{(4j+3)/d_{\psi}}} \right)
\end{aligned}
\end{equation}
\noindent
Here, $\epsilon=10000$ is the assumed maximum value of $x$ and $y$ along the corresponding axes within the WSI. From the above equation, there are four components derived for a given embedded dimension $j$. In order to ensure that the resulting positional encoding vector maintains the same dimensionality as the feature vector $\psi$, we further define $j\in\{0,..,d_{\psi}/4\}$.

In our case, while the width and height of a WSI can reach hundred thousands pixels, in practice, we can normalize the patch locations into relative positioning. For example, by extracting patches of size $512 \times 512$ and stride of $512 \times 512$ from a WSI of $51200 \times 51200$, we can effectively denote each patch belonging to a $100 \times 100$ canvas. The $x$ and $y$ positions of each image patch are thus ensured to be smaller than the $\epsilon$ limit defined in \cref{eq:positional_encoding}.

\subsection{Handcrafted prototypical patterns}

Recently, features based on co-localization of specific nuclei types such as TILs have been shown to be robust and prognostic\cite{Shaban2019TILScore}. In addition, there is recent evidence to suggest that morphology of tissue components can also be predictive \cite{Diao2021HandcraftWSI-Mutation}.

Given the formulation in \cref{eq:MHA}, the importance of encoding positional information and the successes so far of its less general variants on WSI prediction tasks, we first assume that the resulting representation vector from MHA is highly discriminative. In addition, it also contains information on the variation within instance features (or instance-level patterns) and how these variations co-occur with each other (instance-level co-localization patterns). Under this formulation, H2T representation of WSIs offers the following:

\begin{itemize}
\item We can disentangle the features related to instance-level patterns from those for co-localization of patterns;
\item Instead of learning prototypical patterns in a supervised manner, as in some of the recent works, we can provide our own set of reference patterns;
\item Rather than learning the attribution of each instance for composing the WSI representation, we can derive an effective attribution ourselves;
\item Similarly, we can also devise the co-localization in a handcrafted manner.
\end{itemize}

Going forward, we use instance $\psi$ to denote an image patch's feature vector within a WSI. This image patch can be of arbitrary size and from an arbitrary magnification level.

\begin{figure*}[!t]
\centering
\includegraphics[width=1.0\linewidth]{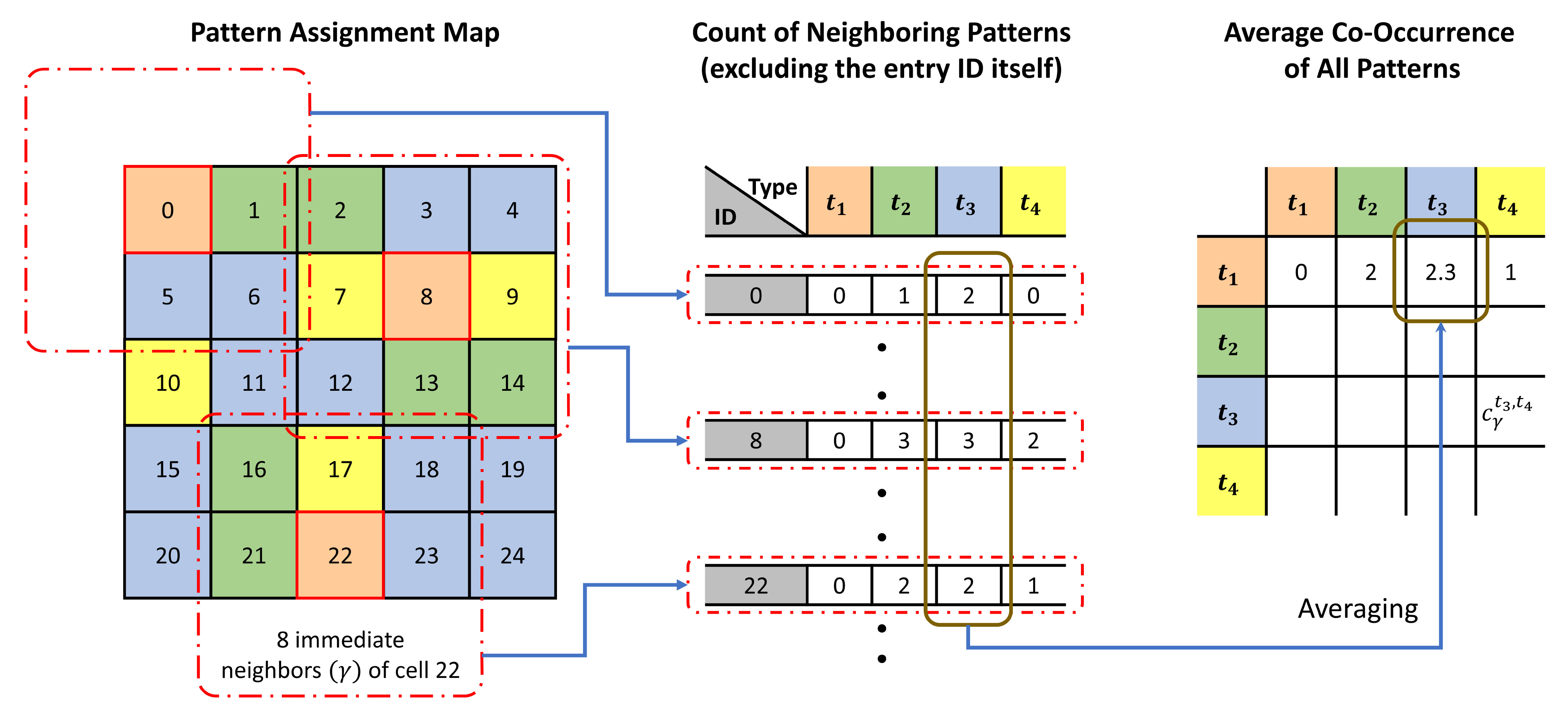}
\caption{Step by step illustration of how \cref{eq:colocalization} is utilized to calculate the pattern co-localization matrix (PCM). Here, each pattern (or type $t_{k}$) is denoted in a different color. It is worth noting that the pattern counting does not include the type of entry (or the central) cell.
}
\label{f:colocalization_matrix}
\end{figure*}

\subsubsection{Representations from histological patterns}

Prototypical patterns of a set of image patches (or strictly speaking, their feature vectors) can be obtained via clustering. While there are many clustering techniques, not many of them scale well when processing millions of input samples, as in in our case. Therefore, given the large amount of image patches and the high-dimensionality of their feature vectors, we use $k$-means clustering. Utilization of $k$-means and nearest neighbors has been noted to be particularly effective for tasks on the same magnitude of difficulties \cite{kilian2019simpleshot-knn}. For this usage, it is crucial to normalize the feature vector of each image patch with $L_2$-norm.

As a result, the (prototypical) histological patterns $p \in P$ are also the resulting centroids obtained from the clustering process. With $\psi$ denoting the feature vector of an image patch, we therefore reformulate \cref{eq:MHA_HopField} into the following form,
\begin{equation}
\label{eq:Hi_formula}
\overline{H_i} = \frac{1}{|\Phi_i|} \sum_{\forall \psi_j \in \Phi_i }{f(p_i, \psi_j) \odot \psi}
\end{equation}
\begin{equation}
\label{eq:H_formula}
\overline{H} = Concat(\overline{H_0}, ..., \overline{H_N})
\end{equation}
\noindent
where $\overline{H_i}$ is representation when projecting the WSI against the $i$-th prototypical histological pattern $p_i$ and $\Phi_i$ is the set of image patches assigned to $p_i$. Specifically, a patch $\psi$ is assigned to a pattern $p_i$ when the distance between their representations is the smallest compared to all other patterns. In \cref{eq:Hi_formula}, $f(p_i, \psi_j)$ is an attribution function that measures the similarity between $p_i$ and $\psi_j$ and $\odot$ denotes the element-wise multiplication of two vectors. The resulting $\overline{H}$ is, therefore, the WSI representation when projected against a derived set of histological prototypical patterns $P$. We refer to  $\overline{H}$ as weighted average pooling (WAP) features in \cref{f:pipeline}.

There are several ways to derive the attribution of each instance $\psi$ (image patch) with respect to their assigned prototypical pattern $p_i$. Assuming that both $p_i$ and $\psi_j$ have already been normalized by $L_2$-norm and $d(p_i, \psi_j)$ is their Euclidean distance, we investigate the following attribution function $f(p_i, \psi_j)$:

\begin{enumerate}
\item $\overline{H}$ : Average pooling of all assigned $\psi_j$, we effectively set $f(p_i, \psi_j)=1$ in this scenario.
\item $\overline{H}$-w : Weighted pooling of the assigned $\psi_j$. Here, the weights are the inverse distance between $p_i$ and its $\psi_j$. Thus, we define $f(p_i, \psi_j)=1 - d(p_i, \psi_j)$.
\item $\overline{H}$-t[X] : Similar to \#1, with $\Phi_i$ further filtered such that only patches having $d(p_i, \psi_j)\geqslant X $ are selected for aggregation.
\item $\overline{H}$-k[X] : Similar to \#1, with $\Phi_i$ further filtered such that only top X instances that are the \textbf{closest} to $p_i$ are selected for aggregation.
\item $\overline{H}$-fk[X] : Similar to \#1, with $\Phi_i$ further filtered such that only top X-th instances that are the \textbf{furthest} to $p_i$ are selected for aggregation.
\end{enumerate}

\subsubsection{Representations from co-localization of patterns}

Inspired by how features describing co-localization of different nuclei types can be constructed \cite{Abbet2020DivideandRuleSL}, we define the pattern co-localization matrix (PCM) as follows,

\begin{equation}
\label{eq:colocalization}
\begin{aligned}
c^{i,j}_\gamma &= \frac{1}{|\Phi^{i,j}_\gamma|}\sum_{\psi \in \Phi^{i,j}_\gamma }{u^{i,j}_\gamma} \\
\widehat{C}(\gamma) &=
    \begin{bNiceMatrix}
    c^{1,1}_\gamma & \Cdots & c^{1,j}_\gamma \\
    \Vdots         & \Ddots & \Vdots         \\
    c^{i,1}_\gamma & \Cdots & c^{i,j}_\gamma \\
    \end{bNiceMatrix}
\end{aligned}
\end{equation}
\noindent
where $\Phi^{i,j}_\gamma$ is a set of patches where each instance not only belongs to pattern $p_i$ but is also surrounded by patches of pattern $p_j$ within the radius $\gamma$. We additionally denote $u^{i,j}_\gamma$ as the number of patches assigned to pattern $p_j$ within the neighborhood of patch $\psi\in \Phi^{i,j}_\gamma$. With these definitions, $c^{i,j}_\gamma$ can be understood as the average occurrence of pattern $p_j$ around pattern $p_i$ within the distance $\gamma$. Meanwhile, $\widehat{C}(\gamma)$ is the average pattern co-localization (PCM) matrix of all patterns within the WSI. Finally, we only study the 8 immediate neighbors in this paper. This calculation is illustrated in \cref{f:colocalization_matrix}.

While we can extend the number of $\gamma$ for assessment and stack many resulting $\widehat{C}(\gamma)$ together for a more detailed representation, they still only reflect one aspect of the co-localization distribution, which is their mean. It would become unscalable when trying to incorporate longer distance and/or other distribution measurements.

To resolve this issue, we can employ CNNs (or graph neural networks for a more general form) to learn the patterns of co-occurrence. Specifically, with a set of patches $\Phi_i$ assigned to each pattern $p_i$ from \cref{eq:Hi_formula}, because we know the position of each patch within the original WSI, we can therefore project such assignments back to their 2D relative positioning. By repeating this process for all prototypical patterns, we obtain an image which we denote as the Pattern Assignment Map (PAM). It is worth noting that this projection is akin to a coarse segmentation process i.e. patch-wise classification rather than pixel-wise classification. It is expected that a neural network trained on this image can therefore learn the the patterns of co-occurrence.

In the case of using CNNs, the prototypical PAM is not the same as a normal image where each pixel value is a category rather than the raw pixel intensity. We are therefore encouraged to encode and train the CNNs on such encoding rather than learning the PAM directly. For categorical values like ours, one-hot encoding is an exceptionally cheap and effective way for such modeling. To differentiate with the handcrafted co-occurrence features from \cref{eq:colocalization}, we term the features obtained from training CNNs as Deep PAM features and denote them as $\overline{C}$.



\begin{table*}[t!]
\caption{{Summary of the main datasets used in our experiments. It is worth noting that the Normal WSIs utilized here are adjacent to the Tumor WSIs within the biopsy samples. Additionally, TCGA-Lung (NSCLC) is a combination of tumorous WSIs within TCGA-LUAD and TCGA-LUSC dataset.}
}
\label{t:dataset_summary}
\centering
\small
\begin{NiceTabular}{llrr|rr}[]
\toprule
\RowStyle{\bfseries}
\Block[c]{}{Datasets} & \Block[c]{}{Tissue Type/Organ} & \Block[c]{}{FFPE WSIs} & \Block[c]{}{Frozen WSIs} & \Block[c]{}{Total WSIs} & Total Patients \\
\midrule
\RowStyle{\bfseries}
\textbf{CPTAC-LUAD} & Lung & 0 & 1048 & 1048 & 111 \\
& \lvl Normal & \lvl 0 & \lvl 374 & \lvl 374 & \lvl - \\
& \lvl Tumor (Adenocarcinoma) & \lvl 0 & \lvl 674 & \lvl 674 & \lvl - \\
\cdashlinelr{1-6}
\RowStyle{\bfseries}
\textbf{CPTAC-LUSC} & Lung & 0 & 1025 & 1025 & 108 \\
& \lvl Normal & \lvl 0 & \lvl 363 & \lvl 363 & \lvl - \\\
& \lvl Tumor (Squamous Cell Carcinoma) & \lvl 0 & \lvl 662 & \lvl 662 & \lvl - \\
\cdashlinelr{1-6}
\RowStyle{\bfseries}
\textbf{TCGA-LUAD} & Lung & 531 & 1067 & 1598 & 522 \\
& \lvl Normal & \lvl 0 & \lvl 244 & \lvl 244 & \lvl - \\
& \lvl Tumor (Adenocarcinoma) & \lvl 531 & \lvl 823 & \lvl 1354 & \lvl - \\
\cdashlinelr{1-6}
\RowStyle{\bfseries}
\textbf{TCGA-LUSC} & Lung & 512 & 1200 & 1712 & 504 \\
& \lvl Normal & \lvl 0 & \lvl 347 & \lvl 347 & \lvl - \\
& \lvl Tumor (Squamous Cell Carcinoma) & \lvl 512 & \lvl 853 & \lvl 1365 & \lvl - \\

\midrule

\RowStyle{\bfseries}
\textbf{TCGA-Lung (NSCLC)} & Lung --- Non-Small Cell Carcinoma & 1043 & 1676 & 2719 & 1026 \\
& \lvl Adenocarcinoma & \lvl 531 & \lvl 823 & \lvl 1354 & \lvl 522 \\
& \lvl Squamous Cell Carcinoma & \lvl 0 & \lvl 662 & \lvl 662 & \lvl 504 \\
\cdashlinelr{1-6}
\RowStyle{\bfseries}
\textbf{TCGA-Breast (BRCA)} & Breast --- Carcinoma & 992 & 1317 & 2309 & 975\\
& \lvl Invasive Ductal  &\lvl 791 &\lvl 1090 &\lvl 1881 &\lvl 776 \\
& \lvl Lobular & \lvl 201 & \lvl 227 & \lvl 428 & \lvl 199 \\
\cdashlinelr{1-6}
\RowStyle{\bfseries}
\textbf{TCGA-Kidney (RCC)} & Kidney --- Renal Cell Carcinoma & 909 & 1441 & 2350 & 926 \\
& \lvl Clear Cell & \lvl 504 & \lvl 1062 & \lvl 1566 & \lvl 523\\
& \lvl Papillary Cell & \lvl 296 & \lvl 236 & \lvl 532 & \lvl  290\\
& \lvl Chromophobe Cell & \lvl 109 & \lvl 143 & \lvl 252& \lvl 113\\
\midrule
\midrule
\RowStyle{\bfseries}
\textbf{Total (Unique)} & Lung, Breast, Kidney & 2944 & 7098 & 10042 & 3048\\
\bottomrule
\end{NiceTabular}
\end{table*}

\section{Experimental Results}

\subsection{Datasets}

{
For this study, we utilized 6 different datasets consisting of a total of 10,042 \textit{unique} WSIs from 3048 \textit{unique} patients from The Cancer Genome Atlas (TCGA) and Clinical Proteomic Tumour Analysis Consortium (CPTAC). The number of WSIs and the distribution of associated labels within each dataset are summarized in \cref{t:dataset_summary}. We constructed TCGA-Lung (NSCLC) dataset by using only tumorous WSIs within TCGA-LUAD and TCGA-LUSC dataset.
}

{
For the same patient, in addition to the tissue slides that contain tumorous area, there are also normal adjacent tissue slides. Thus, for lung tissue, there are 3 WSI-levels: Normal, Lung Adenocarcinoma (LUAD) and Lung Squamous Cell Carcinoma (LUSC). For breast tissue, there are 2 WSI-level labels: Invasive Ductal (IDC) and  Lobular Carcinoma (ILC). Lastly, for kidney, there are 3 WSI-level labels: Clear Cell, Papillary, and Chromophobe Renal Cell Carcinoma (CCRCC, PRCC, CHRCC). Although there are slides that may come from the same patient, for simplicity, in this study we treated each WSI as an independent sample. 
}

Aside from the TCGA and CPTAC cohorts, we also utilized 2 WSIs from the ACDC \cite{li2020acdc} dataset for rough qualitative assessment.

\subsection{Evaluation}

Our H2T framework is a handcrafted approximation of the inner working of the Transformer. Therefore, it is of interest to determine how closely H2T approximates the performance of the original Transformer architecture. In order to assess this, we specifically trained two Transformer models as baseline: transformer-1 with only one multi head attention (MHA) layer for the final aggregation;  and transformer-2 with one multi head self-attention (MHSA) layer and one MHA layer for the final aggregation.

We linearly probe the discriminative power of our resulting WSI-level representation on a series of classification tasks. {This is a widely utilized technique in the computer vision community for assessing feature representation obtained from self-supervised learning \cite{chen2020SimCLR, caron2020SWAV, zhang2016colorful}}. Specifically, the features are considered to be usable only if they are highly discriminative out-of-the-box. In other words, the degree of their discriminative power is reflected by their linear-separability. In our case, the discriminative power of the resulting WSI-level H2T representations directly correlates with the usability of the prototypical patterns utilized to construct them. Finally, this probing is achieved by inputting the features through a single linear layer for making the prediction.

In subsequent experiments, we used either CPTAC or TCGA as the discovery set, namely being training and validation set. On the other hand, we kept the entire other cohort as evaluation set (independent testing set). Within the discovery set, we split the cohort across both labels and the subset (such as CPTAC-LUAD and CPTAC-LUSC) into 5 folds in a stratified manner. For each fold, we then selected the best model and validated it on the testing cohort. Subsequently, we reported the mean and standard deviation obtained from each fold from both the discovery and evaluation cohort.

{We evaluate our representation on several classification tasks: Normal vs Tumor of lung tissue (LUAD and LUSC are combined to make Tumor label), LUAD vs LUSC, Normal vs LUAD vs LUSC of lung tissue, CCRCC vs PRCC vs CHRCC, or IDC vs ILC. To be in line with existing methods that have been applied for the Normal vs Tumor and cancer sub-typing tasks}, we use area under the receiver operating characteristic (AUROC) as the evaluation metric. However, due to the skewed distribution of labels within the dataset, we additionally calculate average precision (AP), which is another way to compute the area under the precision-recall curve (AUPRC), for each label and report their mean (mAP).

Aside from assessing the predictive power of our proposed WSI-level representation, we are also interested in how prototypical patterns which originate from different source tissue (the reference cohort) would impact downstream analysis. In this study, we consider two scenarios: tissue coming from different centres and/or different tissue types. Throughout the main text, we focus on the first scenario: using all available tissue within TCGA or CPTAC when one of them is the discovery cohort. {For example, with LUAD vs LUSC, when TCGA is used as the discovery cohort, all prototypical patterns are extracted using \textit{only} TCGA data whereas CPTAC data are kept intact as independent testing set.} On the other hand, the Supplementary Material explores the latter case: using \textbf{only Normal tissue} within TCGA or CPTAC when one of them is the discovery cohort.

\begin{figure*}[!ht]
\centering
\includegraphics[width=0.95\textwidth]{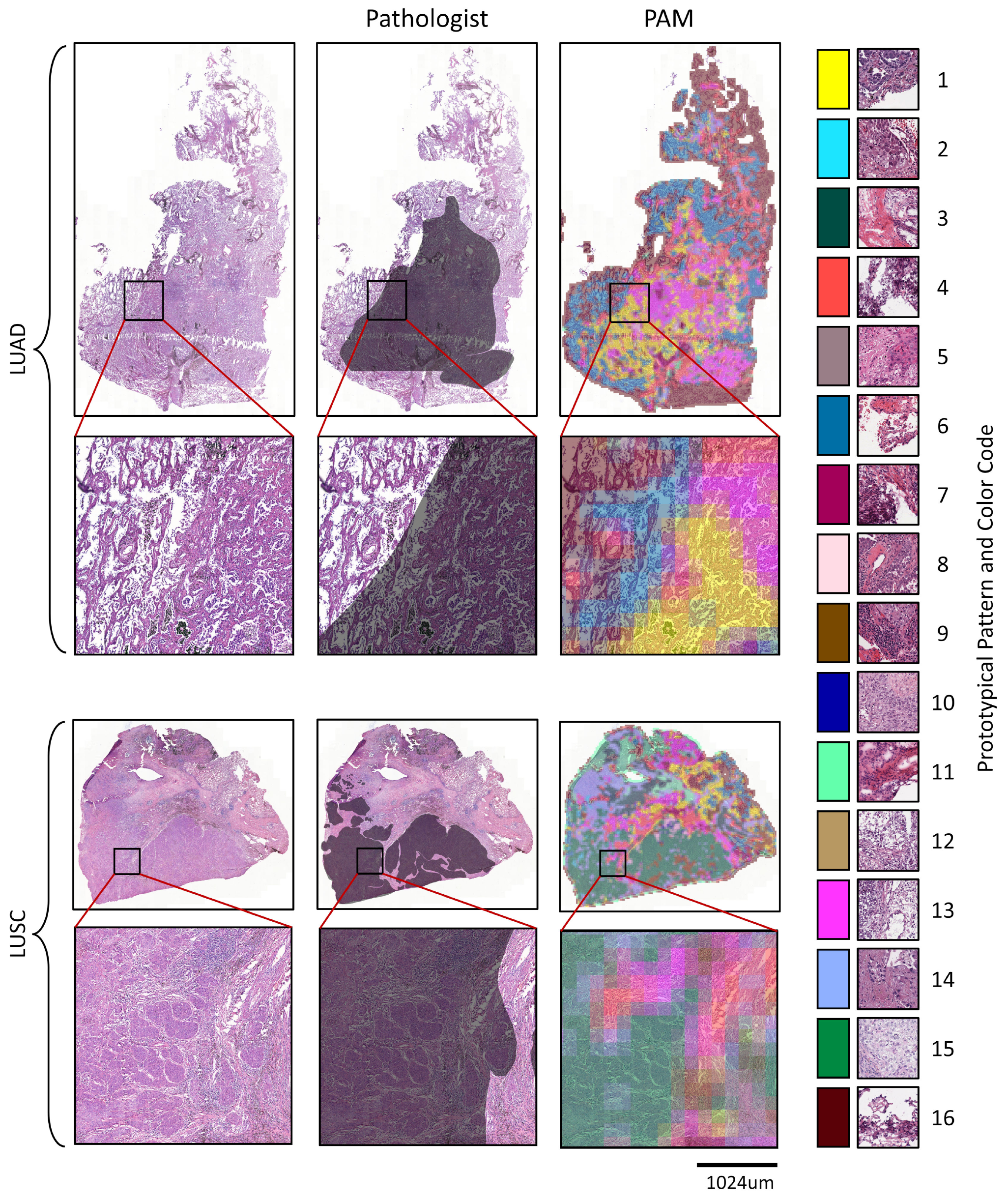}
\caption{Comparison of pattern assignment maps (PAMs) against the annotations from the pathologists {on lung tissue}. PAMs were constructed using 16 prototypical patterns. By using SWAV-ResNet50 patch-level features, these patterns were derived from \textcolor{blue}{all WSIs} within the TCGA cohort. The sample WSIs in the figure are from ACDC cohort and the overlaid areas with dark shade are tumorous regions provided by pathologists. We visually identify that prototypical patterns with ocean blue color (6) or deep red color (16) are closely related to Normal tissue areas; patterns with yellow (1) or pink (13) colors are related to tumorous area in LUAD; and pattern with green color (15) is related to LUSC. When selecting out areas having these tumorous patterns and quantitative measuring them against the pathologist annotations, we obtained 0.6879 ($p\ll0.0001$) and 0.8407 ($p\ll0.0001$) in Pearson correlation coefficient for LUAD and LUSC respectively. The same set of prototypical patterns were later utilized for \cref{f:pam2}.
}
\label{f:pam-pathologists}
\end{figure*}

\begin{figure*}[!ht]
\centering
\includegraphics[width=0.95\textwidth]{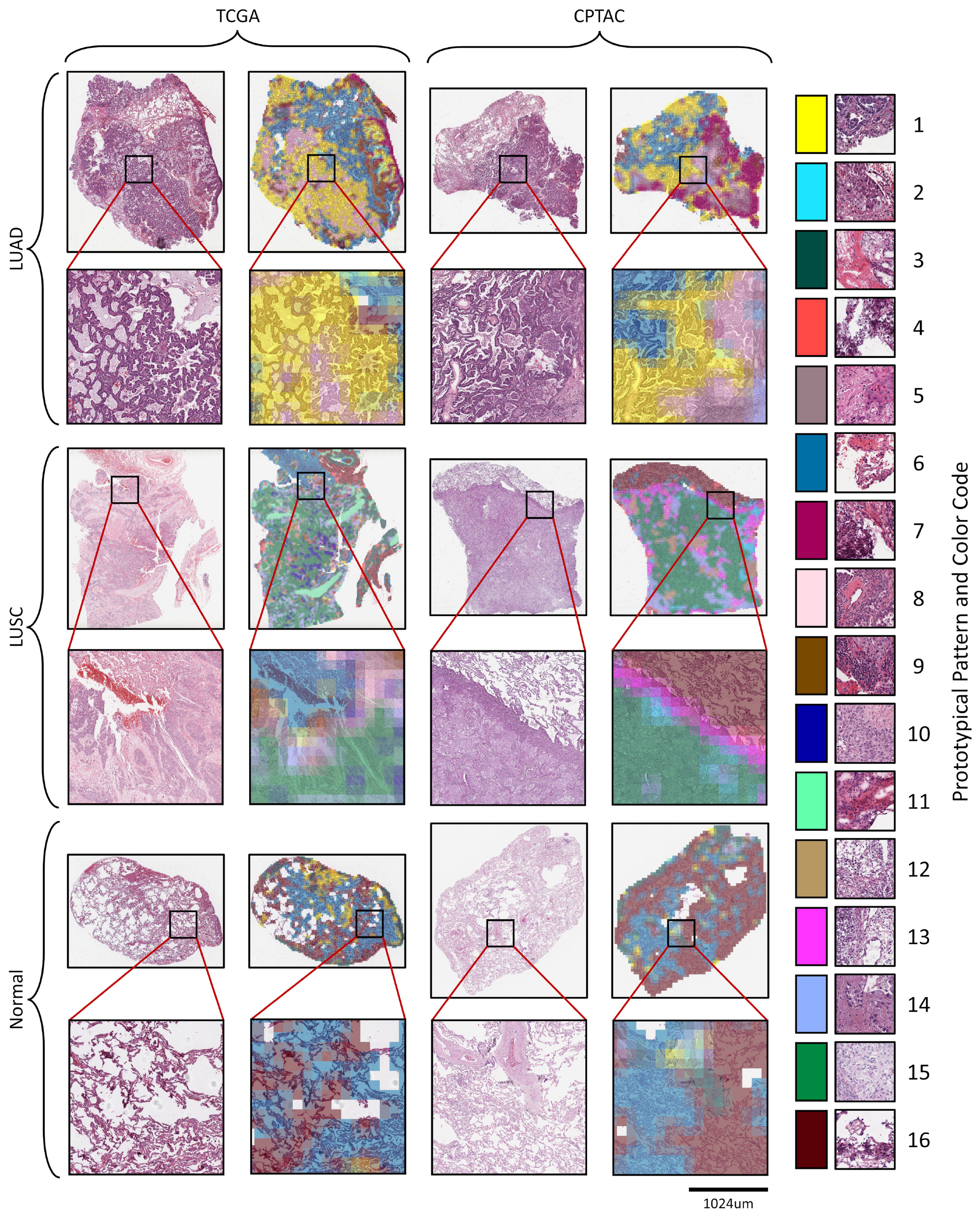}
\caption{Visual comparison of the pattern assignment maps (PAMs) {of lung tissue} between the reference cohort (all WSIs in TCGA) and other cohorts (all WSIs in CPTAC). PAMs were constructed using 16 prototypical patterns. In turn, using SWAV-ResNet50 patch-level features, these patterns were derived based on \textcolor{blue}{all WSIs} within the TCGA cohort. Overlapping regions have their colors averaged for illustration. Locations whose colors do not align with established color code indicate the transition between assigned patterns. Note that patterns having the same color but were derived from different clustering may not be semantically similar. The colors in \cref{f:pam-pathologists} denote the same assignments as this figure. The color assignment is for assessing the consistency within this figure only.
}
\label{f:pam2}
\end{figure*}

\begin{figure*}[!ht]
\centering
\includegraphics[width=1.0\textwidth]{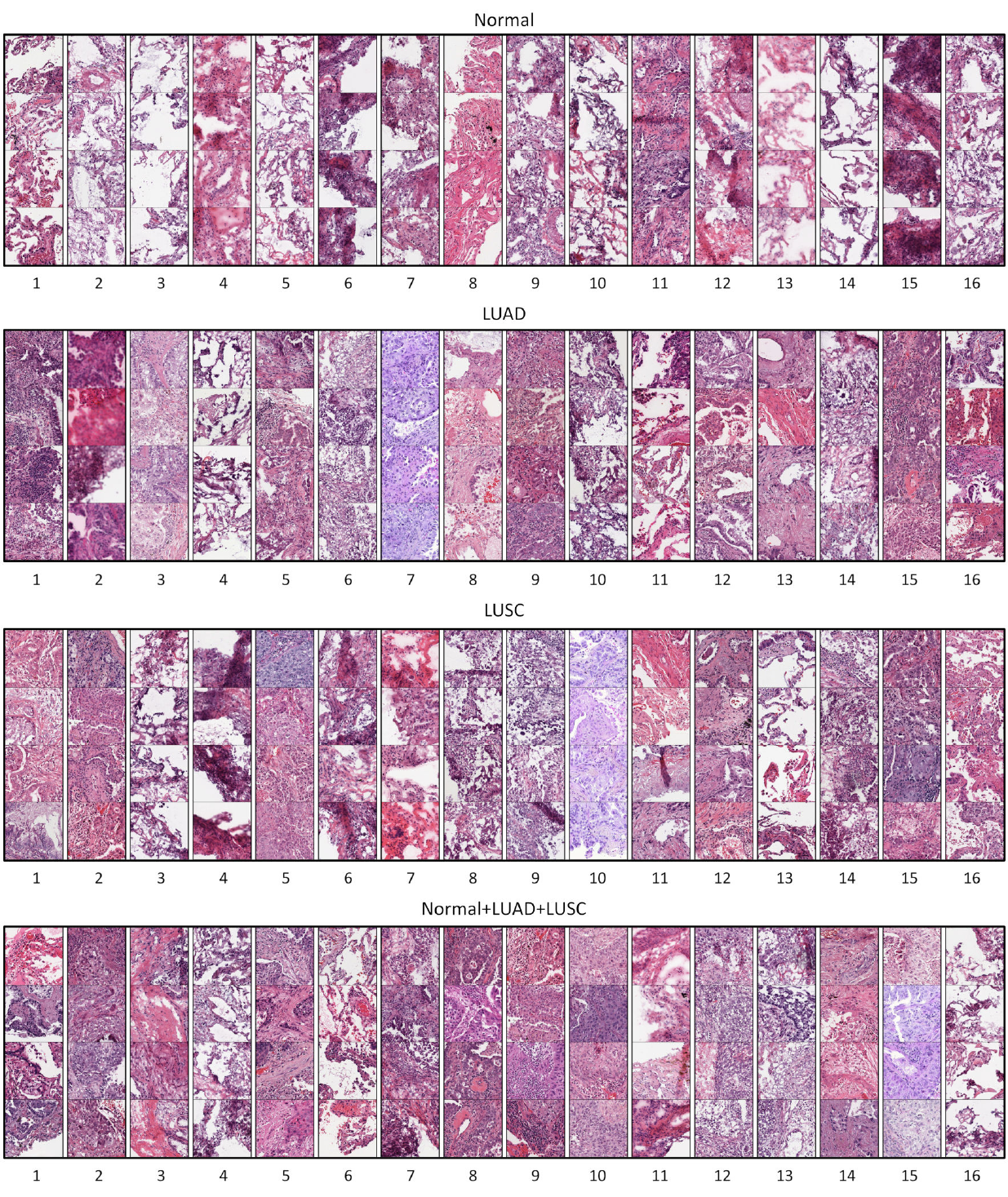}
\caption{{Visualization of prototypical patterns extracted using solely lung tissue}. Four closest patches to each prototypical pattern. There are 16 prototypical patterns derived using SWAV-ResNet50 patch-level features from only Normal WSIs, only LUAD WSIs, only LUSC WSIs or all available WSIs (Normal+LUAD+LUSC) within the TCGA cohort. Within each set of reference tissue (WSIs), the patterns are arranged in a 1$\times$16 (height$\times$width) grid where each cell contains 4 closest patches.
\vspace*{12px}
}
\label{f:patterns_samples}
\end{figure*}

\subsection{Implementation details}

We extracted patches of shape $512 \times 512$ with $256 \times 256$ degree of overlapping out from each WSI. To avoid redundant information, we focused on patches coming mostly from tissue area. Afterward, we applied a pretrained ResNet50 on each patch to derive their representations. Depending on each experimental setup described further below, these patches are either at $0.25$ or $0.50$ micron per pixel (mpp), corresponding to $40\times$ or $20\times$ magnification respectively.

For the Transformer baseline models, we constructed both MHSA and MHA with 8 attention heads. For the aggregation layer in particular, according to \cref{eq:MHA_HopField}, it has $R$ of shape $16 \times 2048$. In other words, in each attention head, there are 16 learnable prototypical patterns each of which is described by a 2048-dimensional feature vector. We trained each Transformer model for 50 epochs using Adam optimizer with a learning rate of 0.001. {We provide more details about their formulation in the Supplementary Material (\cref{s:baseline_transformers}).}

For experiments related to learning the co-localization of prototypical patterns, we utilized a reduced version of ResNet50 with 34 layers. To differentiate this from the usual ResNet34, this version uses full bottleneck (3 consecutive convolutional layers with kernel size of 1, 3 and 1) instead of 2 convolutional layers where each has kernel size of 3. Similarly, we also trained this model for 50 epochs using Adam optimizer with a learning rate of 0.001.

Regarding linear probing, we trained all the linear layers, which are also known as fully connected layers, for 50 epochs using Adam optimizer with a learning rate of 0.001.

Finally, although we compare our proposal against methods that utilized patch-level features obtained from ResNet34 or ResNet18 pretrained on histological images, we utilized only patch-level features based on ImageNet to assess our proposal. There are two primary reasons for this decision: a) models based on ImageNet are widely used and therefore have been extensively assessed, b) recent research has also shown that models pretrained on histological images may not always provide better performance compared to those trained on ImageNet, especially for deeper neural networks like ResNet50 \cite{ciga2020self}.

\subsection{Prototypical patterns}

Before diving deeper into further assessments, it is important to conduct a sanity check on the sets of prototypical patterns we derived from a histological perspective. We constructed 4 sets of prototypical patterns using solely Normal, LUAD, LUSC and all WSIs (Normal+LUAD+LUSC) in the TCGA cohort as reference. Using patch-level representations obtained from SWAV-ResNet50 \cite{caron2020SWAV}, we extracted 16 prototypical patterns for each set.
While the number of prototypical patterns can be different, results from our ablation studies (as reported in \cref{s:ablation_clusters}) have shown that 16 are generally enough to obtain discriminative WSI-level representation.

To validate whether the prototypical patterns and their resulting assignments carry histologically meaningful information, we compare the resulting PAMs against annotations of tumor regions provided by pathologists in \cref{f:pam-pathologists}. Here, the sample WSIs are taken from the ACDC dataset \cite{li2020acdc} while the PAMs were based on patterns obtained when using the entire TCGA lung cohort as the reference cohort. We visually identify that prototypical pattern with ocean blue color (color with code 6) or deep red (16) are closely related to Normal tissue areas. Meanwhile, prototypical patterns with yellow (1) or pink (13) colors are related to tumorous area in LUAD whereas green color (15) is related to LUSC. When considering areas assigned with these tumorous patterns and quantitatively measuring them against the pathologist annotations, we obtained 0.6879 ($p\ll0.0001$) and 0.8407 ($p\ll0.0001$) in Pearson correlation coefficient for LUAD and LUSC, respectively. As a side note, we restrict this quantitative measurement only to the two sample WSIs provided above. We also note that the existing annotations from the pathologists are rough. As evident from the illustration in \cref{f:pam-pathologists}, often they circled an entire area that contains not just tumorous components. Therefore, we only selected one WSI per category that has the best localized annotation. For our purposes, the annotation needs to be more fine-grained and localized. Future work will involve further validation against better sources of ground truth that satisfy such criteria.

Using the same set of prototypical patterns utilized in \cref{f:pam2}, we further examine the PAMs between the reference cohort (TCGA) and the unseen cohort (CPTAC) on 3 tissue types: LUAD, LUSC and Normal. From the illustration, although the assignments in unseen cohorts are less distinct compared to the reference set, we observe that the assignments still maintain their consistency from a bird's eye point of view for major tissue components, such as tumorous or stromal regions across WSIs in each cohort. Aligning with our previous observations in \cref{f:pam-pathologists}, we also observe similar assignments in \cref{f:pam2} for the sample LUAD and LUSC WSIs. Specifically, in both TCGA and CPTAC cohort, we notice that Tumor WSIs contain a large swath of yellow (0) or brown color (14) compared to their normal counterparts. We provide additional examples when using only Normal WSIs as reference set in the Supplementary Materials (\cref{f:pam1}).

Since different source of reference tissue results in different set of prototypical patterns, we additionally evaluate how the histological meaning of these sets vary by examining the closest patches assigned to each pattern in \cref{f:patterns_samples} when using Normal, LUAD, LUSC or all WSIs (Normal+LUAD+LUSC) in the TCGA cohort as the reference cohort. From the figure, we observe that patches assigned to the same prototypical pattern are semantically similar.

In conclusion, through a brief visual assessment and a rough quantitative measurement, we postulate that our derived prototypical patterns are histologically meaningful. However, further research is necessary to accurately validate the biological meaning of these sets of prototypical patterns.

\begin{table*}[!ht]
\caption{Comparison study on classifying \textcolor{blue}{Normal vs Tumor} WSIs {using solely lung tissue}. The proposed H2T representations ($\overline{H}$ and  $\overline{C}$) were derived based on 16 prototypical patterns. These patterns were obtained by using SWAV-ResNet50 patch-level features extracted from all WSIs within each discovery cohort. $\overline{H}$-w is obtained by weighted summing patch features assigned to a pattern; $\overline{H}$-k128 is obtained by averaging features from the top 128 closest patches assigned to a pattern; $\overline{C}$-one-hot is the representation obtained by training CNN on the one-hot-encoded pattern assignment map (PAM); $\widehat{H}$ is the histogram of the patterns within PAM; $\widehat{C}$ is the co-localization matrix of patterns within PAM. Reported results are mean $\pm$ standard deviation of AUROC taken across 5 stratified folds.}
\label{t:comparative_Normal-Tumor}
\centering
\resizebox{0.8\textwidth}{!}{
\begin{NiceTabular}{cc|cc|cc}[]
\toprule
\RowStyle{\bfseries}
Features & Method & CPTAC-valid & TCGA-test & TCGA-valid & CPTAC-test \\
\midrule
Tuned\tabularnote{Pretrained CNN was further tuned before being used as a feature extractor by the authors in the original work.} & MS-MIL-RNN\tabularnote{Results are taken from Table 2 in \cite{Li2021DSMIL}. {The original work did not report the standard deviation, thus we set it to 0.}} \cite{Fuchs2019Nature}  & - & -    & 0.956±0.000 & -          \\
Tuned\textsuperscript{\textit{a}} & MS-ABMIL\textsuperscript{\textit{b}} \cite{Noriaki2020MS-ABMIL} & - & -    & 0.979±0.000 & -          \\
Tuned\textsuperscript{\textit{a}} & DSMIL-LC\textsuperscript{\textit{b}} \cite{Li2021DSMIL} & - & -    & 0.982±0.000 & -          \\
\midrule
\Block{3-1}{SUPERVISE-ResNet50 \\ \cite{he2016deep}} & CLAM\tabularnote{Results are produced by us.} \cite{lu2020CLAM} & 0.992±0.003 & 0.955±0.007 & {\color[HTML]{0000FF} 0.997±0.002} & {\color[HTML]{0000FF} 0.979±0.002} \\
& transformer-1 & 0.987±0.009 & 0.953±0.009 & 0.992±0.002 & 0.958±0.005 \\
& transformer-2 & {\color[HTML]{0000FF} 0.993±0.005} & {\color[HTML]{0000FF} 0.963±0.004} & 0.996±0.001 & 0.970±0.003 \\
\cdashlinelr{1-6}
\Block{3-1}{SWAV-ResNet50 \\  \cite{caron2020SWAV}} & CLAM\textsuperscript{\textit{c}} \cite{lu2020CLAM} & 0.994±0.005 & 0.970±0.003 & 0.995±0.002 & 0.971±0.004 \\
& transformer-1 & {\color[HTML]{0000FF} 0.993±0.003} & 0.972±0.009 & 0.996±0.002 & 0.961±0.006 \\
& transformer-2 & 0.992±0.004 & {\color[HTML]{0000FF} 0.975±0.002} & {\color[HTML]{0000FF} 0.996±0.003} & {\color[HTML]{0000FF} 0.976±0.004} \\
\midrule
\Block{7-1}{SWAV-ResNet50 \\  \cite{caron2020SWAV}} & $\widehat{H}$ & 0.950±0.004 & 0.880±0.003 & 0.946±0.006 & 0.817±0.010 \\
& $\widehat{C}$ & 0.972±0.002 & 0.939±0.003 & 0.965±0.004 & 0.906±0.008 \\
& $\widehat{H}$+$\widehat{C}$ & 0.979±0.005 & 0.935±0.002 & 0.967±0.007 & 0.872±0.009 \\
& $\overline{H}$-w & 0.996±0.003 & 0.975±0.003 & 0.997±0.001 & 0.954±0.005 \\
& $\overline{H}$-k128 & {\color[HTML]{0000FF} 0.998±0.002} & {\color[HTML]{0000FF} 0.984±0.003} & {\color[HTML]{0000FF} 0.996±0.004} & {\color[HTML]{0000FF} 0.978±0.006} \\
& $\overline{C}$-one-hot & 0.970±0.010 & 0.937±0.007 & 0.966±0.007 & 0.923±0.017 \\
& $\overline{H}$-w+$\overline{C}$-onehot & 0.995±0.004 & 0.972±0.007 & 0.997±0.002 & 0.957±0.007 \\
\bottomrule
\end{NiceTabular}
}
\end{table*}

\subsection{Comparative evaluation}
\label{h:comparative_study}

\textbf{Settings.} It is expected that our proposed set of representations should perform at least comparable to methods that are capable of utilizing all patch-level features from the constituent parts (instances) within the WSI, especially in comparison to the baseline Transformers. In order to evaluate this, we restrict our comparison to other multiple instance learning methods that do not involve majority voting of instance predictions within the WSI: MS-MIL-RNN \cite{Fuchs2019Nature}, CLAM \cite{lu2020CLAM}, MS-ABMIL \cite{Noriaki2020MS-ABMIL}, DSMIL-LC \cite{Li2021DSMIL}, FocAttn-MIL \cite{kalra2021PAF}, {HIPT \cite{chen2022hipt}}  and our two baselines Transformers as described above.

We compared these methods with our proposed representations that were derived by weighted pooling features of patches assigned to each pattern ($\overline{H}$-w), average pooling features of the top 128 closest patches assigned to each pattern ($\overline{H}$-k128), learned co-localization of one-hot-encoded PAM ($\overline{C}$-one-hot) or their combination. In addition to that, we also extracted traditional features: proportion of assigned prototypical patterns $\widehat{H}$ within PAM, the co-occurrence matrix $\widehat{C}$ from \cref{eq:colocalization} or their combination as additional baselines. For this experiment, we constructed these representations from 16 prototypical patterns when using patch-level representations from SWAV-ResNet50.

\textbf{Results.} We respectively present our results for Normal vs Tumor classification and for LUAD vs LUSC using solely lung tissue in \cref{t:comparative_Normal-Tumor} and in \cref{t:comparative_LUAD-LUSC}. Here, the ``Features'' column in both tables denotes the encoders for the patch-level representation: SUPERVISE-ResNet50, SWAV-ResNet50 or fine-tuned/retrained a CNN (Tuned). For the last category, this was often performed on pathological dataset rather than ImageNet \cite{Li2021DSMIL, kalra2021PAF}. For WSI-level H2T representation, for simplicity, we only constructed them using patch-level feature from SWAV-ResNet50. In this experiment, CPTAC-LUAD and CPTAC-LUSC were combined to make the CPTAC dataset whereas TCGA-LUAD and TCGA-LUSC were combined to make the TCGA dataset.

In general, we observe that the Transformer models perform better than all recently published methods on both classification tasks. Furthermore, a full Transformer model (transformer-2) is more powerful than its simplified counterpart (transformer-1). Other than that, using better patch-level representation often results in better performance. This is evident with the model achieving the best performance, transformer-2. When moving from SUPERVISE-ResNet50 to SWAV-ResNet50, on average, its AUROC values for Normal vs Tumor were respectively improved by $1.2\%$  (0.963 vs 0.975) for TCGA-test and by $0.63\%$ (0.970 vs 0.976) for CPTAC-test. In case of LUAD vs LUSC, the improvement is $4.7\%$ (0.796 vs 0.843) for TCGA-test and $1.2\%$  (0.911 vs 0.922) for CPTAC-test.

Interestingly, CLAM achieved comparable or slightly better results in comparison to the transformer-2 in some cases. When using SUPERVISE-ResNet50 for Normal vs Tumor, CLAM achieved higher AUROC on average compared to transformer-2 (0.979 vs 0.970) for CPTAC-test. In comparison to transformer-1, CLAM outperformed the model by $0.2\%$  (0.955 vs 0.953) in AUROC on average for TCGA-test. However, when using SWAV-ResNet50, CLAM performance was slightly worse than the transformer-1 on average by $0.3\%$  (0.972 vs 0.970) in AUROC for TCGA-test. Similar phenomena can also be observed for LUAD vs LUSC. As shown in \cref{t:comparative_LUAD-LUSC}, when using SWAV-ResNet50, transformer-1 outperformed CLAM by $1.0\%$ (0.928 vs 0.918) in AUROC for CPTAC-test. This discrepancy in CLAM performance when switching the origin of patch-level features can be attributed to the fact that CLAM was designed to tune the patch-level representations. As such, when the features are already highly discriminative in case of SWAV, their proposed loss would reduce the representation power instead.

\begin{table*}[!t]
\caption{Comparison study on classifying \textcolor{blue}{LUAD vs LUSC} WSIs. The proposed H2T representations ($\overline{H}$ and  $\overline{C}$) were derived based on 16 prototypical patterns. These patterns were obtained by using SWAV-ResNet50 patch-level features extracted from all WSIs within each discovery cohort. $\overline{H}$-w is obtained by weighted summing patch features assigned to a pattern; $\overline{H}$-k128 is obtained by averaging features from the top 128 closest patches assigned to a pattern; $\overline{C}$-one-hot is the representation obtained by training CNN on the one-hot-encoded pattern assignment map (PAM); $\widehat{H}$ is the histogram of the patterns within PAM; $\widehat{C}$ is the co-localization matrix of patterns within PAM. Reported results are mean $\pm$ standard deviation of AUROC taken across 5 stratified folds.}
\label{t:comparative_LUAD-LUSC}
\centering
\resizebox{0.8\textwidth}{!}{
\begin{NiceTabular}{cc|cc|cc}[]
\toprule
\RowStyle{\bfseries}
Features & Method & CPTAC-valid & TCGA-test & TCGA-valid & CPTAC-test \\
\midrule
Tuned\tabularnote{Pretrained CNN was further tuned before being used as a feature extractor by the authors in the original work. {The original work did not report the standard deviation, thus we set it to 0.}}  & {FocAttn-MIL \cite{kalra2021PAF}}  & - & - & 0.920±0.000 & - \\
\midrule
\Block{3-1}{SUPERVISE-ResNet50 \\ \cite{he2016deep}} & CLAM\tabularnote{Results are produced by us.} \cite{lu2020CLAM} & 0.967±0.004 & 0.791±0.008 & 0.927±0.009 & 0.907±0.007 \\
& transformer-1 & 0.960±0.009 & 0.780±0.005 & 0.912±0.006 & 0.901±0.009 \\
& transformer-2 & {\color[HTML]{0000FF} 0.978±0.005} & {\color[HTML]{0000FF} 0.796±0.005} & {\color[HTML]{0000FF} 0.927±0.008} & {\color[HTML]{0000FF} 0.911±0.008} \\
\cdashlinelr{1-6}
\Block{3-1}{SWAV-ResNet50 \\  \cite{caron2020SWAV}} & CLAM\textsuperscript{\textit{b}} \cite{lu2020CLAM} & 0.977±0.005 & 0.840±0.003 & 0.938±0.007 & 0.918±0.003 \\
& transformer-1 & 0.979±0.007 & 0.835±0.008 & 0.937±0.011 & {\color[HTML]{0000FF} 0.928±0.006} \\
& transformer-2 & {\color[HTML]{0000FF} 0.983±0.008} & {\color[HTML]{0000FF} 0.843±0.005} & {\color[HTML]{0000FF} 0.943±0.012} & 0.922±0.003 \\
\midrule
\Block{7-1}{SWAV-ResNet50 \\  \cite{caron2020SWAV}} & $\widehat{H}$ & 0.745±0.025 & 0.579±0.001 & 0.661±0.016 & 0.770±0.005 \\
& $\widehat{C}$ & 0.859±0.016 & 0.638±0.003 & 0.732±0.020 & 0.791±0.004 \\
& $\widehat{H}$+$\widehat{C}$ & 0.861±0.013 & 0.638±0.003 & 0.728±0.022 & 0.794±0.005 \\
& $\overline{H}$-w & 0.972±0.009 & 0.788±0.013 & 0.927±0.016 & 0.903±0.007 \\
& $\overline{H}$-k128 & {\color[HTML]{0000FF} 0.984±0.004} & {\color[HTML]{0000FF} 0.802±0.005} & {\color[HTML]{0000FF} 0.943±0.010} & {\color[HTML]{0000FF} 0.924±0.005} \\
& $\overline{C}$-onehot & 0.863±0.030 & 0.628±0.013 & 0.704±0.031 & 0.765±0.011 \\
& $\overline{H}$-w+$\overline{C}$-onehot & 0.983±0.006 & 0.789±0.009 & 0.919±0.009 & 0.904±0.006 \\
\bottomrule
\end{NiceTabular}
}
\end{table*}

Regarding our proposed representations, for Normal vs Tumor, when cross-validating within TCGA cohort, $\overline{H}$-w, $\overline{H}$-k128 and $\overline{H}$-w+$\overline{C}$-one-hot based on SWAV-ResNet50 achieved more than 0.99 in AUROC and surpassed DSMIL-LC (0.982). Amongst them, $\overline{H}$-k128 is the most discriminative representation. When being independently tested, $\overline{H}$-k128 in particular achieved better performance compared to the best model on TCGA-test (transformer-2 based on SWAV-ResNet50) by $0.9\%$ (0.984 vs 0.975); and comparable results compared to the best model on CPTAC-test (CLAM based on SUPERVISE-ResNet50).

On LUAD vs LUSC, $\overline{H}$-k128 achieved comparable performance in comparison to the best model (transformer-2) when cross-validating. However, it is noticeably worse than other approaches that involves more complicated neural networks. Most notably on TCGA-test, $\overline{H}$-k128 based on SWAV-ResNet50 achieved only 0.8021 in AUROC in comparison to the worst model under the same setup (0.835 in AUROC of transformer-1 with SWAV-ResNet50).

Meanwhile, although traditional features like $\widehat{H}$, $\widehat{C}$ and $\widehat{H}$+$\widehat{C}$ could achieve results above 0.90 AUROC in some cases in cross-validation for Normal vs Tumor, they severely lack discriminative power on unseen cohort and are out-competed by other methods. On the other hand, despite showing better results compared to handcrafted features in some situations, in comparison to $\overline{H}$-w and $\overline{H}$-k128, representations obtained from learning co-localization ($\overline{C}$) performed poorly when being used alone in both tasks.

{
We further evaluated the performance of our proposed method when sub-typing cancers for other tissue types from the same center. We performed CCRCC vs PRCC vs CHRCC on kidney tissue using RCC dataset and IDC vs ILC on breast tissue using BRCA dataset. We present their results in \cref{t:extended_comparative}. For completeness, we also provide results for LUAD vs LUSC (i.e. using NSCLC dataset). Aside from HIPT, all other methods utilized SWAV-ResNet50 features. Among comparative MIL methods, transformer-2 remains the best performing model. We also observe that our proposed methods remain competitive across different tissue types. In RCC, $\overline{H}$-w achieved similar performance as transformer-2 for AUROC but it has higher mAP. Compared to HIPT, $\overline{H}$-w AUROC is $0.2\%$ higher. Similarly, for BRCA, $\overline{H}$-w outperformed transformer-2 by $0.3\%$ and is significantly better than HIPT by $6.5\%$.
}

With these results, we demonstrate that the representations from H2T can achieve results comparable to other state-of-the-art approaches. {To further understand the difference in performance of these methods, we also performed statistical analysis, the results and details are provided in the Supplementary Section ( \cref{s:statistical_analysis}).}

\begin{table*}[!t]
\caption{WSI-level cancer sub-typing for kidney (RCC), breast (BRCA), and lung (NSCLC). Reported results are mean $\pm$ standard deviation of AUROC and mAP taken across 5 stratified folds. For RCC sub-typing, we report the macro-averaged AUROC across the three subtypes. The results of NSCLC are partially copied from column TCGA-valid in \cref{t:comparative_LUAD-LUSC}. Here, CLAM, transformer-1, transformer-2 and our proposed methods utilized \textbf{SWAV-ResNet50} features.}
\label{t:extended_comparative}
\centering

\resizebox{0.8\textwidth}{!}{
\begin{NiceTabular}{c|cc|cc|cc}[]
\RowStyle{\bfseries}
 & \Block{1-2}{RCC} & & \Block{1-2}{BRCA} & & \Block{1-2}{NSCLC} &  \\
\midrule
\RowStyle{\bfseries}
Method & AUROC & mAP & AUROC & mAP & AUROC & mAP \\
\midrule
HIPT\tabularnote{Results are taken from the original paper.} \cite{chen2022hipt} & 0.980±0.013 & - & 0.874±0.060 & - & 0.952±0.021 & - \\
\cdashlinelr{1-7}
CLAM\tabularnote{Results are produced by us.} \cite{lu2020CLAM} & 0.990±0.001 & 0.972±0.003 & 0.926±0.012 & 0.891±0.017 & 0.938±0.007 & 0.936±0.007 \\
transformer-1 & 0.990±0.001 & 0.972±0.007 & 0.928±0.012 & 0.891±0.018 & 0.937±0.011 & 0.936±0.012 \\
transformer-2 & {\color[HTML]{0000FF} 0.993±0.002} & {\color[HTML]{0000FF} 0.981±0.005} & {\color[HTML]{0000FF} 0.936±0.011} & {\color[HTML]{0000FF} 0.905±0.016} & {\color[HTML]{0000FF} 0.943±0.012} & {\color[HTML]{0000FF} 0.944±0.012} \\
\cdashlinelr{1-7}
$\widehat{C}$ & 0.926±0.012 & 0.771±0.027 & 0.749±0.023 & 0.668±0.021 & 0.732±0.020 & 0.723±0.022 \\
$\widehat{H}$ & 0.864±0.008 & 0.677±0.023 & 0.722±0.023 & 0.635±0.020 & 0.660±0.015 & 0.647±0.012 \\
$\widehat{H}$+$\widehat{C}$ & 0.927±0.009 & 0.783±0.029 & 0.768±0.014 & 0.680±0.017 & 0.728±0.022 & 0.717±0.024 \\
$\overline{H}$-k128 & 0.993±0.002 & 0.983±0.005 & 0.924±0.008 & 0.879±0.018 & {\color[HTML]{0000FF} 0.943±0.010} & {\color[HTML]{0000FF} 0.941±0.011} \\
$\overline{H}$-w & {\color[HTML]{0000FF} 0.993±0.002} & {\color[HTML]{0000FF} 0.983±0.003} & {\color[HTML]{0000FF} 0.939±0.005} & {\color[HTML]{0000FF} 0.899±0.014} & 0.927±0.016 & 0.926±0.017 \\
\bottomrule
\end{NiceTabular}
}
\end{table*}

\subsection{Ablation study}

Going along the framework in \cref{f:pipeline}, we investigate various components that are involved in the derivation of the prototypical patterns and the construction of the subsequent WSI-level H2T representation. {Further experiments were conducted and described in \cref{s:extended_ablation}}.

\subsubsection{Representations from co-localization}

\textbf{Settings.} The co-localization of clinically-grounded patterns like TILs has shown to be important for clinical settings. Additionally, our results in \cref{t:comparative_Normal-Tumor} and \cref{t:comparative_LUAD-LUSC} suggest that such co-localization features still remain somewhat predictive even when we compute them based on the abstract patterns rather than the clinically-grounded entities (nucleus types). Here, we investigate further on:

\begin{enumerate}[label=\alph*.]
\item The effect of using one-hot encoding to aid the training process.
\item How the representations learned by CNN fare against their handcrafted counterpart across different sources of prototypical patterns.
\end{enumerate}

The evaluation was conducted by classifying Normal vs Tumor. For this experiment, the prototypical patterns were obtained by using SWAV-ResNet50 features and from all WSIs within the discovery cohort. Accordingly, we compare the discriminative power of: $\widehat{C}$ the co-localization matrix of patterns within the pattern assignment map (PAM); $\overline{C}$-raw the representation obtained by training CNN on the PAM; $\overline{C}$-one-hot the representation obtained by training CNN on the one-hot-encoded PAM.

\textbf{Results.} From the results in \cref{t:ablation_colocalization_B}, we identify that one-hot encoding is critical for training CNNs when using PAMs as input. Regardless of the number of prototypical patterns, without one-hot encoding, the performance would drop up to 0.14 in mAP. Interestingly, for this particular task, representations learned by CNNs performed noticeably better than handcrafted features. However, when using 16 prototypical patterns, the latter performed comparable to the former. This suggests the results may vary for a more difficult task. We provide additional study related to this within the Supplementary Material (\cref{t:ablation_colocalization_Normal-LUAD-LUSC}) which further highlights this possibility. 

\begin{table*}[t!]
\caption{Ablation study on WSI-level H2T representations based on the co-localization of prototypical patterns by classifying \textcolor{blue}{Normal vs Tumor} {using solely lung tissue}. Using patch-level features from SWAV-ResNet50, all WSIs (Normal+LUAD+LUSC) within the discovery cohort were utilized to derive each set of prototypical patterns. $\overline{C}$-raw is the representation obtained by training CNN on the pattern assignment map (PAM); $\overline{C}$-one-hot is the representation obtained by training CNN on the one-hot-encoded PAM; $\widehat{C}$ is the co-localization matrix of patterns within PAM. Reported results are mean $\pm$ standard deviation of mAP taken across 5 stratified folds.}
\label{t:ablation_colocalization_B}
\centering
\resizebox{0.65\textwidth}{!}{
\begin{NiceTabular}{cc|cccc}[]
\toprule
\RowStyle{\bfseries}
{\#} of clusters & Method & CPTAC-valid & TCGA-test & TCGA-valid & CPTAC-test \\
\midrule
8 & $\widehat{C}$ & {\color[HTML]{0000FF} 0.968±0.002} & 0.809±0.002 & {\color[HTML]{0000FF} 0.912±0.008} & 0.882±0.007 \\
8 & $\overline{C}$-raw & 0.933±0.019 & 0.764±0.010 & 0.805±0.036 & 0.909±0.006 \\
8 & $\overline{C}$-one-hot & 0.902±0.008 & {\color[HTML]{0000FF} 0.952±0.027} & 0.867±0.018 & {\color[HTML]{0000FF} 0.916±0.012} \\
\cdashlinelr{1-6}
16 & $\widehat{C}$ & 0.971±0.002 & 0.851±0.004 & 0.915±0.019 & 0.897±0.008 \\
16 & $\overline{C}$-raw & 0.934±0.017 & 0.744±0.021 & 0.799±0.029 & 0.826±0.006 \\
16 & $\overline{C}$-one-hot & {\color[HTML]{0000FF} 0.971±0.010} & {\color[HTML]{0000FF} 0.865±0.009} & {\color[HTML]{0000FF} 0.928±0.013} & {\color[HTML]{0000FF} 0.926±0.016} \\
\cdashlinelr{1-6}
32 & $\widehat{C}$ & {\color[HTML]{0000FF} 0.977±0.003} & 0.826±0.007 & 0.924±0.025 & 0.884±0.005 \\
32 & $\overline{C}$-raw & 0.826±0.087 & 0.667±0.024 & 0.732±0.043 & 0.862±0.014 \\
32 & $\overline{C}$-one-hot & 0.966±0.012 & {\color[HTML]{0000FF} 0.836±0.017} & {\color[HTML]{0000FF} 0.943±0.009} & {\color[HTML]{0000FF} 0.906±0.010} \\
\bottomrule
\end{NiceTabular}
}
\end{table*}

\subsubsection{Pooling strategies}

\begin{table*}[t!]
\caption{Ablation study on using different pooling strategies for \cref{eq:H_formula} constructing WSI-level H2T representations. The task is classifying \textcolor{blue}{Normal vs LUAD vs LUSC} {using solely lung tissue}. All available (Normal+LUAD+LUSC) WSIs within the discovery cohort were utilized to derive 16 prototypical patterns. Reported results are mean $\pm$ standard deviation of mAP taken across 5 stratified folds.}
\label{t:ablation_pooling}
\centering
\resizebox{\linewidth}{!}{
\begin{NiceTabular}{c|cc|cc|cc|cc}[]
\RowStyle{\bfseries}
& \Block{1-4}{SWAV-ResNet50} & & & & \Block{1-4}{SUPERVISE-ResNet50} & & & \\
\toprule

\RowStyle{\bfseries}
Method & CPTAC-valid & TCGA-test & TCGA-valid & CPTAC-test & CPTAC-valid & TCGA-test & TCGA-valid & CPTAC-test \\


\midrule
$\overline{H}$ & 0.974±0.005 & 0.806±0.008 & 0.939±0.015 & 0.845±0.006 & 0.961±0.006 & 0.756±0.004 & 0.900±0.011 & 0.809±0.008 \\
$\overline{H}$-w & {\color[HTML]{0000FF} 0.977±0.002} & {\color[HTML]{0000FF} 0.809±0.008} & {\color[HTML]{0000FF} 0.942±0.015} & {\color[HTML]{0000FF} 0.858±0.006} & {\color[HTML]{0000FF} 0.964±0.007} & {\color[HTML]{0000FF} 0.762±0.007} & {\color[HTML]{0000FF} 0.906±0.008} & {\color[HTML]{0000FF} 0.817±0.009} \\
\cdashlinelr{1-9}
$\overline{H}$-t0.2 & {\color[HTML]{0000FF} 0.974±0.005} & {\color[HTML]{0000FF} 0.806±0.008} & {\color[HTML]{0000FF} 0.939±0.015} & {\color[HTML]{0000FF} 0.845±0.006} & 0.961±0.006 & 0.756±0.004 & 0.900±0.011 & {\color[HTML]{0000FF} 0.809±0.008} \\
$\overline{H}$-t0.3 & 0.974±0.006 & 0.805±0.008 & 0.938±0.016 & 0.844±0.006 & {\color[HTML]{0000FF} 0.962±0.005} & {\color[HTML]{0000FF} 0.759±0.004} & {\color[HTML]{0000FF} 0.902±0.010} & 0.806±0.006 \\
$\overline{H}$-t0.4 & 0.968±0.007 & 0.793±0.013 & 0.918±0.011 & 0.820±0.011 & 0.933±0.010 & 0.727±0.007 & 0.857±0.010 & 0.747±0.004 \\
$\overline{H}$-t0.5 & 0.864±0.014 & 0.685±0.011 & 0.752±0.023 & 0.680±0.008 & 0.807±0.008 & 0.543±0.014 & 0.692±0.020 & 0.575±0.006 \\
$\overline{H}$-t0.6 & 0.591±0.030 & 0.429±0.004 & 0.476±0.022 & 0.501±0.006 & 0.570±0.008 & 0.413±0.005 & 0.452±0.024 & 0.463±0.005 \\
$\overline{H}$-t0.7 & 0.395±0.006 & 0.369±0.002 & 0.368±0.008 & 0.375±0.001 & 0.433±0.022 & 0.366±0.003 & 0.353±0.005 & 0.390±0.004 \\
\cdashlinelr{1-9}
$\overline{H}$-k8 & 0.978±0.006 & 0.813±0.010 & 0.954±0.012 & 0.871±0.007 & 0.960±0.009 & 0.743±0.004 & 0.917±0.009 & 0.805±0.012 \\
$\overline{H}$-k16 & 0.979±0.005 & 0.820±0.008 & 0.953±0.016 & 0.883±0.008 & 0.968±0.008 & 0.752±0.003 & 0.917±0.019 & 0.821±0.014 \\
$\overline{H}$-k32 & 0.982±0.005 & 0.819±0.006 & 0.955±0.016 & 0.892±0.008 & 0.971±0.008 & 0.763±0.004 & 0.929±0.021 & 0.839±0.015 \\
$\overline{H}$-k64 & {\color[HTML]{0000FF} 0.985±0.004} & 0.825±0.006 & 0.956±0.017 & {\color[HTML]{0000FF} 0.893±0.010} & 0.978±0.006 & 0.769±0.004 & 0.932±0.020 & 0.848±0.015 \\
$\overline{H}$-k128 & 0.984±0.004 & {\color[HTML]{0000FF} 0.825±0.005} & {\color[HTML]{0000FF} 0.957±0.014} & 0.890±0.007 & {\color[HTML]{0000FF} 0.979±0.005} & {\color[HTML]{0000FF} 0.775±0.001} & {\color[HTML]{0000FF} 0.933±0.019} & {\color[HTML]{0000FF} 0.855±0.012} \\
\cdashlinelr{1-9}
$\overline{H}$-fk8 & 0.902±0.015 & 0.672±0.006 & 0.832±0.017 & 0.626±0.012 & 0.873±0.016 & 0.611±0.010 & 0.807±0.024 & 0.684±0.009 \\
$\overline{H}$-fk16 & 0.917±0.015 & 0.690±0.008 & 0.845±0.018 & 0.642±0.013 & 0.893±0.007 & 0.621±0.009 & 0.821±0.023 & 0.705±0.012 \\
$\overline{H}$-fk32 & 0.929±0.012 & 0.713±0.005 & 0.863±0.016 & 0.661±0.013 & 0.902±0.016 & 0.641±0.006 & 0.852±0.027 & 0.725±0.012 \\
$\overline{H}$-fk64 & 0.933±0.013 & 0.725±0.007 & 0.871±0.012 & 0.688±0.015 & 0.917±0.015 & 0.662±0.007 & 0.867±0.030 & 0.744±0.007 \\
$\overline{H}$-fk128 & {\color[HTML]{0000FF} 0.946±0.010} & {\color[HTML]{0000FF} 0.732±0.010} & {\color[HTML]{0000FF} 0.887±0.012} & {\color[HTML]{0000FF} 0.719±0.015} & {\color[HTML]{0000FF} 0.930±0.014} & {\color[HTML]{0000FF} 0.685±0.010} & {\color[HTML]{0000FF} 0.893±0.010} & {\color[HTML]{0000FF} 0.760±0.007} \\
\bottomrule
\end{NiceTabular}
}
\end{table*}

\textbf{Settings.} Despite their weaknesses on higher resolution (\cref{t:ablation_features_magnification}), results thus far indicate that our handcrafted formulation in \cref{eq:H_formula} and the H2T framework can be remarkably competitive compared to the Transformer. We further investigate several different ways to derive the set of weights for \cref{eq:H_formula}. For this experiment, we constructed the WSI-level representation using 16 prototypical patterns and from patch-level features extracted from using either SWAV-ResNet50 or SUPERVISE-ResNet50.

\textbf{Results.} Our results are provided in \cref{t:ablation_pooling}. We observe that the source of prototypical patterns remains the utmost important aspect for H2T and can result in a significant difference in performance. In particular, excluding $\overline{H}$-fk, representations based on SUPERVISE-ResNet50 consistently performs worse than those based on SWAV-ResNet50.

Other than that, given a set of patches assigned to a prototypical pattern, based on the results of $\overline{H}$-t[X] (where `[X]' is the threshold value), we identify that selecting patches with distances to their assigned prototypical pattern larger than or equal to `[X]' offer no noticeable improvement in performance. In particular, 
when the threshold is 0.2 ($\text{`[X]'}=0.2$ or $\overline{H}$-t0.2), the performance is equal to that of $\overline{H}$, which has no filtering, across all categories. This suggests that no patches having distances smaller than 0.2. In addition to that, when the thresholds for selection are larger than 0.2, the discriminative power of the resulting WSI-level representation rapidly degrades in comparison to $\overline{H}$.

The results of $\overline{H}$-t[X] suggest that patches within a certain distance to their assigned prototypical patterns may be beneficial for the WSI-level representation. This possibility becomes evident from the results of $\overline{H}$-k[X]. For this set of WSI-level representation, rather than selecting all patches within a certain threshold, we select the `[X]'-th closest patches to its assigned prototypical pattern. We identify that even when using only the 8 closest patches to each pattern, which corresponds to a maximum of 128 patches in total per WSI, it can offer noticeable improvement in comparison to the generic $\overline{H}$ and $\overline{H}$-w. Furthermore, aligning with previous observations in \cref{t:comparative_LUAD-LUSC} and \cref{t:comparative_Normal-Tumor}, selecting the top 32, 64 or 128 patches per pattern provide the most optimal performance. Accordingly, this corresponds to selecting from 512 to 4096 patches at maximum per WSI. As a side note, the maximum here is the theoretical limit because not all patterns have patches assigned to them. This situation has been partially illustrated in \cref{f:pam2} and \cref{f:pam-pathologists}. Therefore, the actual number of selected patches may be less than the theoretical limits.

In contrast to selecting the top closest strategy, selecting the top furthest patch is detrimental to the representation discriminative power in general in comparison to other method. Interestingly, the best $\overline{H}$-fk[X] (with $\text{`[X]'}=128$) was able to achieved high cross-validation results. In case like using SUPERVISE-ResNet50, its TCGA-valid result can approach that of $\overline{H}$.

On a more reserved note, although $\overline{H}$-w does not offer the same performance as selection-based methods, they maintain a relatively good performance out-of-the-box compared to others while having no tuning parameters.

\subsection{Discovery experiments}

Now that we have verified that the features from H2T are discriminative enough for downstream analysis, we provide a brief demonstration on how they can be used for other tasks, such as discovering anomalous WSIs. 

In this experiment, we first considered TCGA as discovery cohort and CPTAC as independent (evaluation) cohort. Furthermore, we assumed to only know about the Normal WSIs within TCGA. Similar to what we have done so far, we started by deriving the prototypical patterns based on all WSIs that we have access to (the entire TCGA lung cohort). Afterward, we generated the $\overline{H}$-w WSIs level representation for all WSIs in both TCGA and CPTAC dataset. Subsequently, we train isolation forest \cite{liu2012IsolationForest}, a simple machine learning method to score anomaly, on these Normal WSI-level representations for scoring all WSI-level representation in both datasets. Anomalous WSIs (or out of distribution) have their scores lower than those considered to be in distribution.

\begin{figure*}[t!]
\centering
\includegraphics[width=1.0\linewidth]{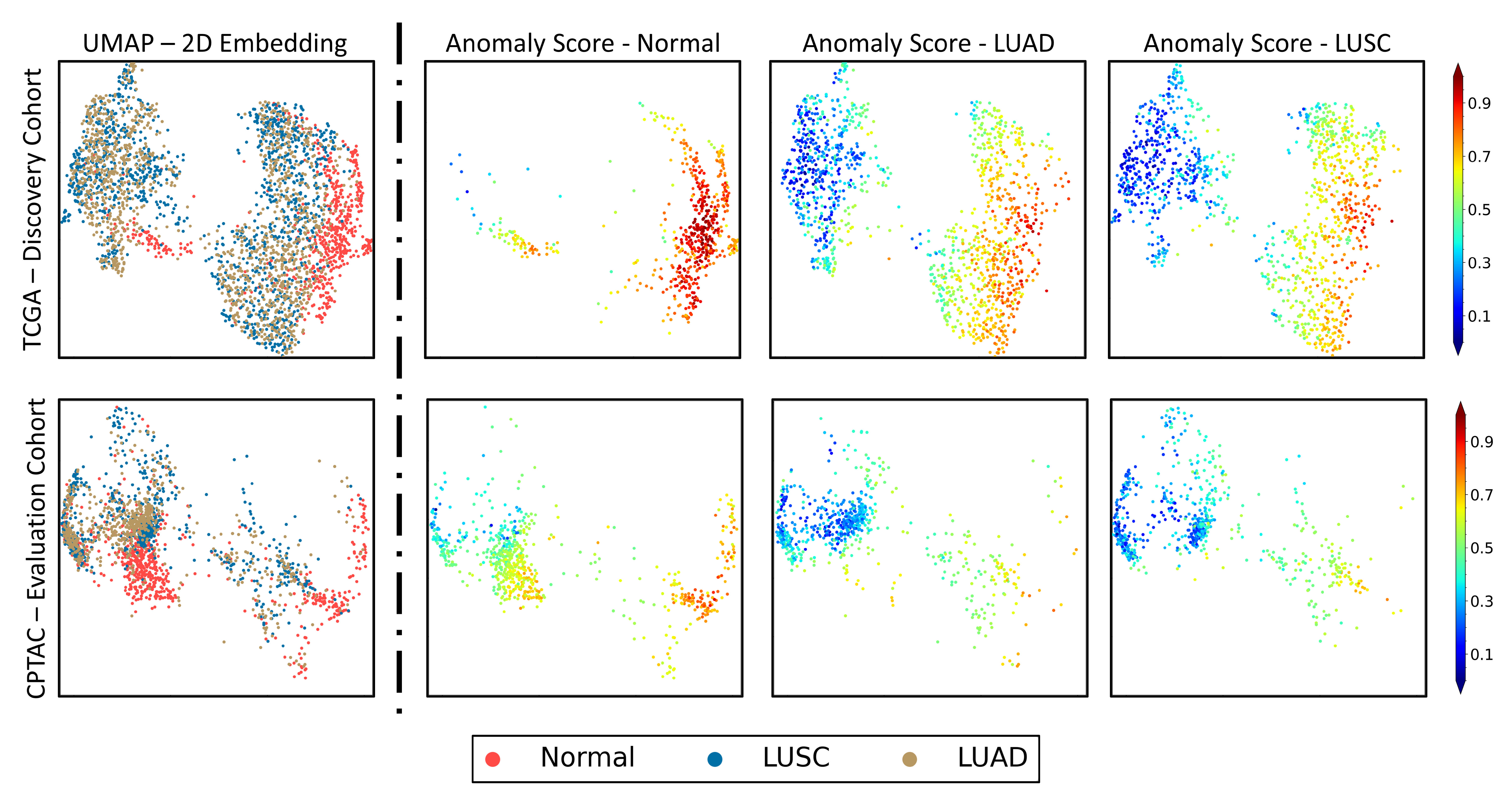}
\caption{{Discovery study using solely lung tissue}. WSI-level H2T representations are projected onto a 2D plane for exploration. The representation was computed based on the prototypical patterns obtained from \textcolor{blue}{Normal} WSIs (using SWAV-ResNet50 patch-level features) within TCGA (discovery cohort). Here, each data point represents 1 WSI. The projection was done using UMAP, which was also trained by using only WSIs within TCGA. UMAP plots show the sample placements and their labels. Meanwhile, the other plots show the out of distribution (anomaly) score assigned to each sample. The lower the score is, the higher the chance the sample is out of distribution (not Normal).
}
\label{f:discovery_ood}
\end{figure*}

The results are shown in \cref{f:discovery_ood}. Because the WSI-level features we derived are high-dimensional (a matrix of $16\times 2048$ at the very least), for visualization purpose, we utilize UMAP \cite{mcinnes2018umap} to project these 32768 features down to 2D plane for positioning each WSI. From the TCGA plot, although there are LUAD and LUSC WSIs which have their anomaly scores in orange range (around 0.8), Normal WSIs still have their anomaly scores distinctly higher (dark deep red or above 0.9). Despite that, there is a number of Normal WSIs having anomaly score below 0.7. However, their positions in the figure are noticeably different from the LUAD and LUSC WSIs, this is reflected by a cluster of red dot on the left within the Embedding subplot. As for the CPTAC subplot, the Normal WSIs have their anomaly scores clearly lower compared to those in the TCGA. most of them have less than 0.8 anomaly score and their scores often concentrate around 0.5 ranges. In spite of that, their scores are still noticeably higher than the majority of LUAD and LUSC WSIs which concentrate below 0.3 spectrum. In addition to the anomaly score, the positions of the Normal WSIs are also distinguishable compared to the cluster of Tumor WSIs.

Other than the anomaly score, the unsupervised clustering of the WSIs within \cref{f:discovery_ood} also shows that unsupervised separation of LUAD and LUSC remains difficult. In TCGA, although the positions of LUAD and LUSC WSIs separate into small clusters, these clusters highly intermixed. On CPTAC, LUAD and LUSC WSIs seem to occupy the same space.

All in all, these observations indicate that while our proposed representation can be used for discovery process, more research into improving its discriminative power is necessary.

\subsection{Runtime complexity}

Given the data-hungry nature of deep learning and the increasingly large amount of data that we have to deal with, it is desirable for a method to be computationally cheap as much as possible while still being strongly predictive. We have touched upon the impact of reducing the time for preparing H2T framework (by using less epochs for clustering) in \cref{f:ablation_clustering_epochs}. Here, we further provide an estimate on the runtime complexity for the Transformer models and our methods in \cref{t:runtime_complexity}. We first emphasize that these numbers would vary depending on the systems and should only be taken as reference. The measurements in \cref{t:runtime_complexity} were made using an NVIDIA-A100 GPU when there were no other processes running. All the methods considered here utilized patches at $mpp=0.50$ and TCGA as discovery cohort. The reported time for training is the average time needed for training 1 single fold split of the TCGA containing an average of 2,560 WSIs. The reported time for feature extraction is the average time taken to finish extracting 1 WSI within the TCGA. On the other hand, the clustering and projection time of H2T are shown for the entire TCGA lung dataset consisting of 3,210 WSIs.

Consistent with the reports in \cite{Vaswani2017Transformer}, Transformer models are notoriously memory demanding. From \cref{t:runtime_complexity}, it is clear that utilizing a full Transformer (transformer-2) or a deeper Transformer (by stacking more multi head self-attention layers) is not possible for common workstation systems. In addition to that, even with a 80GB A100 GPU and a batch size of 1, it is not possible for us to process many WSIs at $mpp=0.25$.

In contrast, even when accounting for the clustering times, our H2T is still much cheaper computationally. When processing a single fold (a fold within the stratified split of TCGA discovery cohort), H2T could be 2.6 to 3.6 times faster compared to the Transformer models. Furthermore, due to its small footprint on GPU memory, by running multiple processes in parallel on the same GPU, it is possible to finish the entire training for TCGA as discovery cohort even faster. In particular, with its small footprint, we can fit 4 to 5 running processes within 12GB of GPU memory, thus bringing the overall differences in processing speed to 10 or 14 times depending on the system. Nonetheless, just like other methods based on patch-level features, feature extraction remains the most time-consuming step.

\begin{table}[t!]
\caption{Runtime complexity of the proposed H2T representation based model and Transformer models. Note that the values here are for reference only as they vary across systems. For our case, the entire experiments were conducted on a single NVIDIA-A100 GPU when there were no other running processes. All the methods under measurement utilized patches at 0.5 micron per pixel and TCGA as discovery cohort. The reported time for training is the average time needed for training 1 single fold split of the TCGA (containing an average of 2,560 WSIs). The reported time for feature extraction is the average time taken to finish extracting 1 WSI within the TCGA. On the other hand, the clustering and projection time of H2T were taken for the entire TCGA dataset.}
\label{t:runtime_complexity}
\centering
\resizebox{1.0\linewidth}{!}{
\begin{tabular}{c|c|ccc}
\toprule
\textbf{Metrics} & \textbf{Steps} & \textbf{transformer-1} & \textbf{transformer-2} & \textbf{H2T} \\
\midrule
\multirow{4}{*}{Time} & Features Extraction & 2min/WSI & 2min/WSI & 2min/WSI \\
& Clustering & - & - & 7min \\
& Projection & - & - & 3min \\
& Training & 110min/fold & 80min/fold & 14min/fold \\
\midrule
\multirow{2}{*}{GPU} & Batch Size & 8 & 4 & 32 \\
& Memory & 40GB & 130GB & 2GB \\
\bottomrule
\end{tabular}
}
\end{table}

\section{Concluding remarks}

Downstream analysis of histopathology images relies on efficient and effective representation of whole slide images (WSIs). In this paper, we proposed a new approach named Handcrafted Histological Transformer (H2T) for deriving holistic WSI-level representations. We have demonstrated that our derived H2T representations can be readily utilized in both supervised and unsupervised manners with relative ease. In the former setting, we have demonstrated that our set of H2T representations are just as predictive as those obtained from the current best methods, namely the Transformer models. In addition to its effectiveness in representing WSIs, the proposed H2T framework is also more computationally efficient. To the best of our knowledge, H2T is the first handcrafted framework that can compete with the Transformer family while requiring less computational resources.

In general, machine learning systems also have trouble adapting to data coming from different distributions, the so-called {\em out of distribution or OOD} problem. In computational pathology, this can be particularly challenging. Owing to staining and data acquisition practices that vary from center to center, models trained on one center thus may not be directly applicable to data from other centers. It is still an open question as to how we can reliably detect the OOD samples and use them to re-calibrate the system under clinical settings. Nonetheless, despite being more automated and reproducible compared to pathologists, these approaches are still far from being a fully automated system that can discover and stratify diseases. We have also shown how anomaly discovery can be made using the H2T representations in an unsupervised manner. At its core, our method is a handcrafted interpretation on how a black-box Transformer architecture actually performs, thus providing better transparency on the decision-making process of the model. Through the creation and subsequent usage of prototypical patterns, it is also possible to further utilize our established clinical knowledge rather than simply abstract patterns mined from the dataset. We hypothesize that prototypical patterns obtained from clustering a set of representative patches from pathologists can be as effective or may be even more predictive compared to using all available patches. As the WSI-level H2T representation is in a way a projection of a WSI against the established knowledge (i.e., the prototypical patterns that we extracted), it may be used to explore how the patterns evolve along the progression of a disease.

{ In this paper, prototypical patterns are extracted from an initial dataset and therefore bounded to a single dataset. A set of prototypical patterns of lung tissue is meaningless for subtyping cancer in breast tissue. Even in the same tissue, a set of prototypical patterns extracted from a subset of lung disease surely would not reflect another. One would assume simply redoing the process using a larger dataset is enough to amend the problem. However, it is not realistic to have \textit{data for all} diseases. In addition, the computation cost of such operation would increase exponentially each time the dataset gets expanded. To alleviate this problem, it is important to investigate how a set of prototypical patterns from one dataset can be considered as ``novel" compared to those extracted using another dataset as well as how we can combine these sets of prototypical patterns together.
}

{
Parallel to the above, further investigations on how to automatically and systematically obtain better prototypical patterns is also important. In the scope of this work, we investigated the plausibility of our proposal extensively by using k-means due to its simplicity. There are better clustering techniques that are available. Dictionary learning is another promising research direction given how the prototypical patterns are obtained and utilized. Lastly, given the close relationship between Transformer-based method and CBIR system, investigating on the potential use of H2T as an approximation of a subset of Transformers for a CBIR task may open new research directions in the nascent area of computational pathology.

Given the robustness and representation power of massive Transformer models like GPT-3, it is also of interest to pretrain a Transformer model end-to-end for computational pathology. However, the massive scale of WSIs at high magnification level presents a huge technical challenge for such an attempt. In light of this challenge, we consider an intermediate compressed representation of a WSI (such as the H2T representation) or multi-stage pretraining as in \cite{chen2022hipt} as two possible novel directions for end-to-end training for WSI-level analysis.
}

Through the H2T framework, we have also shown how WSI-level representation can be disentangled into instance-level patterns and co-localization of instance-level patterns. This disentanglement thus allows us to explore how each representation contributes toward the overall predictive power of the final WSI-level representation. Given the reduction in predictive power of co-localization representation in some tasks, we hypothesize that a prototypical pattern that is a combination of both patch co-localization and patch-level features can be a strong alternative for a better WSI-level representation.

{While we have successfully derived representations at WSI-level and were able to use them to identify anomalous WSIs, identifying which patches in each WSI that contribute in turning the WSI into an anomaly remains difficult. Because these patches potentially indicate new tissue phenotype, without a way to go from WSI-level back to patch-level, explainable prediction and automated identification of disease may not be possible as we still rely on pathologists for reviewing all possible anomalous WSI cases.} In addition, so far we have only demonstrated anomalous detection on a very narrow and easy scope of categories, namely Normal vs Tumor in lung tissue. Our method is not yet able to reliably highlight the cancer subtypes. Further research is required to further enhance its discriminative power. 

Meanwhile, although our framework can provide more transparency compared to other methods, our prototypical patterns are still much more abstract compared to known patterns, such as the number of TILs. Within this study, while we have provided a high-level assessment of the possible histological meaning associated with our prototypical patterns, this was conducted only on a small number of WSIs. Therefore, in order to utilize our prototypical patterns and their WSI-level representation for more clinically related tasks, it is necessary to investigate their histological meaning on a larger scale.

We have uncovered and further confirmed existing practice in machine learning, namely keeping an independent testing set is extremely important to correctly assess the results. Specifically, in the scope of our dataset and tasks, despite each cohort containing thousands of WSIs, we have shown that high cross-validation results within each cohort may not be enough to identify good approaches. This is evident through our ablation study on pooling strategies. Here, in spite of having high performance on cross-validation, many of our models fall short on the evaluation set.

Self-supervised learning in natural images has relied on the complexity of ImageNet to measure progress in the field. Much of the complexity associated with ImageNet can be partly attributed to its size (millions of images) and the large number of present categories, that are structured in a meaningful hierarchy. In comparison, although the largest publicly available dataset in computational pathology also contains thousands of WSIs (TCGA), there is less variation between its categories. As a result, we believe curating a new dataset may be necessary to further develop and investigate WSI-level representations in computational pathology. Specifically, this dataset should be a large collection of extensively stratified diseases (such as cancer grades of all tissue types).

Finally, unlike image patch classification, localization of disease sites as well as anomaly or novelty detection are intrinsically intertwined for processing WSIs. Future work will involve performing localization in an unsupervised manner and further validation on more disease variations.

\section*{Acknowledgments}

We thank Rob Jewsbury and Simon Graham for their invaluable feedback during the write-up of the manuscript. Quoc Dang Vu is funded by The Royal Marsden NHS Foundation Trust. NR and SR are part of the PathLAKE digital pathology consortium, which is partly funded from the Data to Early Diagnosis and Precision Medicine strand of the governments Industrial Strategy Challenge Fund, managed and delivered by UK Research and Innovation (UKRI). NR and SR are also funded by the European Research Council (funding call H2020 IMI2-RIA). NR was also supported by the UK Medical Research Council (grant award MR/P015476/1), Royal Society Wolfson Merit Award and the Alan Turing Institute.

\bibliographystyle{IEEEtran}

\bibliography{refs}

\newpage
\clearpage
\appendix

\setcounter{subsection}{0}
\renewcommand{\thesubsection}{A\arabic{subsection}}
\setcounter{figure}{0}
\renewcommand{\thefigure}{A\arabic{figure}}
\setcounter{table}{0}
\renewcommand{\thetable}{A\arabic{table}}
\renewcommand{\theequation}{A\arabic{equation}}

\begin{figure*}[!ht]
\centering
\includegraphics[width=1.0\textwidth]{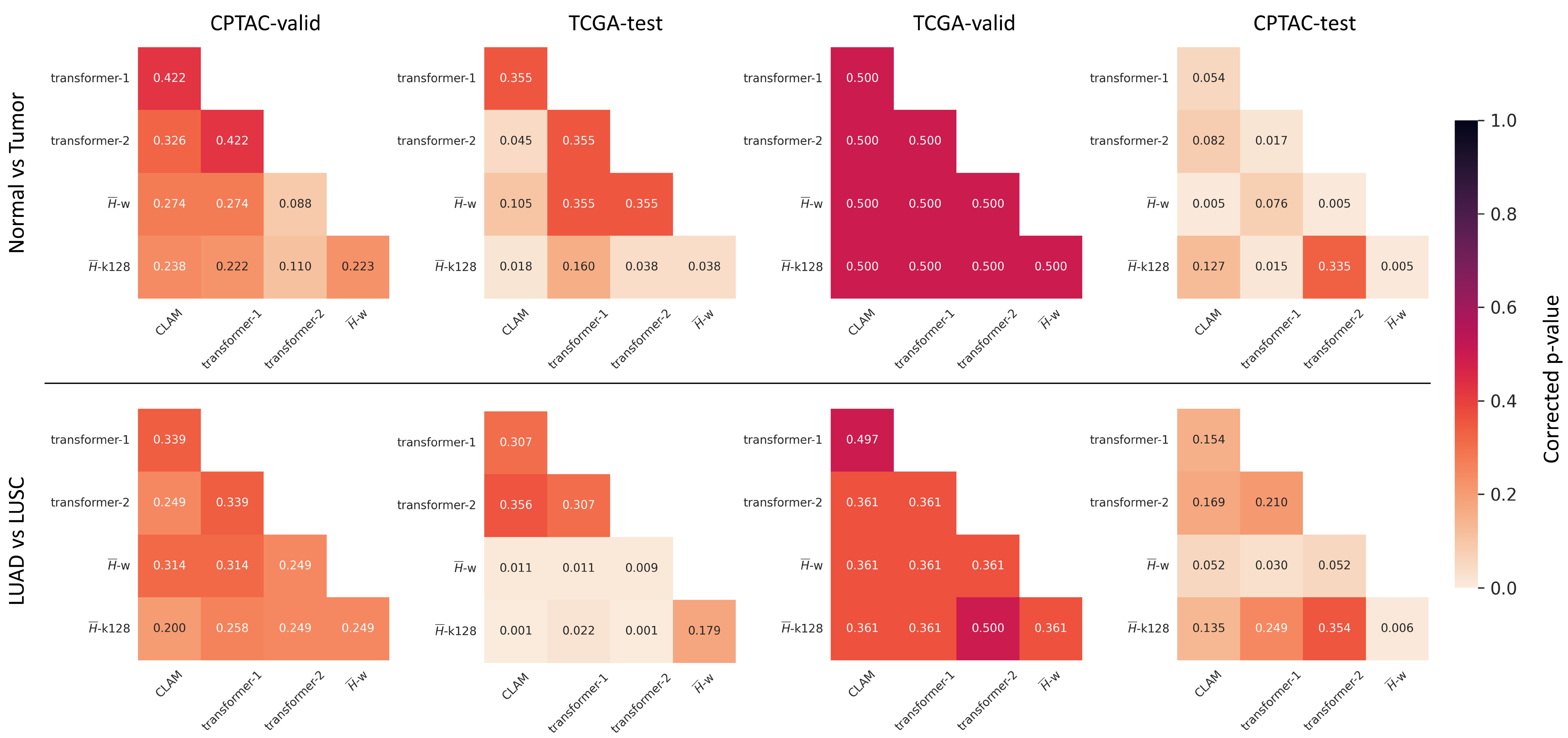}
\caption{$p$ values when doing right-tailed pairwise t-tests on results of best performing models which are reported in \cref{t:comparative_Normal-Tumor} and  \cref{t:comparative_LUAD-LUSC}. The $p$ values were corrected using Benjamini/Hochberg method.
\vspace*{24px}
}
\label{f:p-val-1}
\end{figure*}

\begin{figure*}[!ht]
\centering
\includegraphics[width=1.0\textwidth]{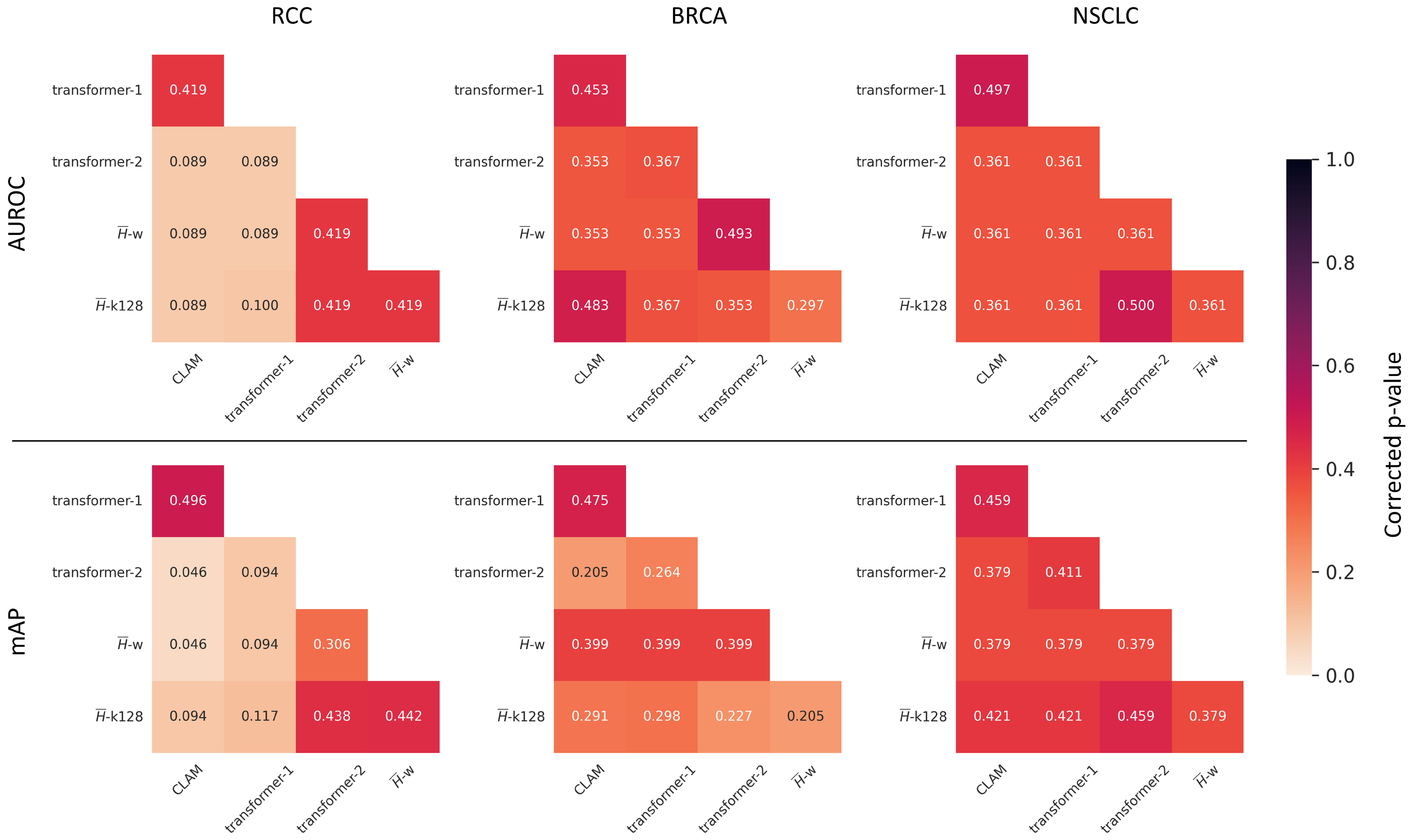}
\caption{$p$ values when doing pairwise right-tailed t-tests on results of best performing models which are reported in \cref{t:extended_comparative}. The $p$ values were corrected using Benjamini/Hochberg method.
\vspace*{24px}
}
\label{f:p-val-2}
\end{figure*}

\section*{Supplementary Material}

{

\subsection{Baseline Transformer architectures}\label{s:baseline_transformers}

Formally, we first define
\begin{itemize}
\item $l$ denotes the $l$-th layer in the network.
\item $h$ denotes the $h$-th head within each MHA layer.
\item $X=\{x_0,...,x_N\}$ is an WSI with $x$ as a feature vector of an image patch that is encoded with positional information.
\item $X^l=\{x^l_0,...,x^l_N\}$ is an WSI representation obtained after the $l$-th layer in the network.
\item $R$ denotes a trainable matrix which is synonymous with prototypical patterns that are learned over the course of the training.
\item $W^l_{Q,h}$ denotes the trainable weight matrix for the query of $h$-th head in the $l$-th layer.
\item $W^l_{K,h}$ denotes the trainable weight matrix for the key of $h$-th head in the $l$-th layer.
\item $W^l_{V,h}$ denotes the trainable weight matrix for the value of $h$-th head in the $l$-th layer.
\item $W^l_C$ denotes the trainable weight matrix for combining the output of all attention heads within a MHA layer.
\item $FCN$ denotes the final classification layer.
\end{itemize}

Based on \cref{eq:MHA} and \cref{eq:MHA_HopField}, The transformer-1 output is defined as follows:
\begin{equation}
\begin{aligned}
output  &=  FCN(X^0) \\
X^0   &= Concat(X^0_0, ..., X^0_H)W^0_c \\
X^0_h &= softmax(
    \beta R W^0_{R,h} {W^0_{K,h}}^T {X}^T
) X W^0_{V,h} \\
\end{aligned}
\end{equation}

On the other hand, transformer-2 output is defined as:
\begin{equation}
\begin{aligned}
output  &=  FCN(X^1) \\
X^1   &= Concat(X^1_0, ..., X^1_H)W^1_c \\
X^1_h &= softmax(
    \beta R W^1_{R,h} {W^1_{K,h}}^T {X^0}^T
) X^0 W^1_{V,h} \\
X^0   &= Concat(X^0_0, ..., X^0_H)W^0_c \\
X^0_h &= softmax(
    \beta X W^0_{R,h} {W^0_{K,h}}^T {X}^T
) X W^0_{V,h} \\
\end{aligned}
\end{equation}

From these formulations, we intuitively see that $R$ in transformer-1 would not be able to represent strong co-localization feature compared to transformer-2. Specifically, $R$ in transformer-1 lacks the full self-attention on all input instances. On the other hand, $R$ in transformer-2 is learned on-top of instance features that are expressed in term of other instances. 

Our empirical results have demonstrated that transformer-2 is better than transformer-1 and they align with our intuition above. However transformer-2 is still quite limited compared to known Transformer models, such as GPT-3. As seen in these massive models, stacking more MHA layers would likely further increase the model capacity in representing co-localization information.

H2T is a handcrafted approximation for transformer-1 but it has no positional information encoded within each patch (i.e instance). Thus, as stated in the main text, it is of our interests to explore how much co-localization information contribute to the baseline performance as we progressively increase the theoretical expression power for co-localization.
}

{{\subsection{Statistical Analysis}}\label{s:statistical_analysis}
As demonstrated in \cref{t:comparative_Normal-Tumor}, \cref{t:comparative_LUAD-LUSC} and \cref{t:extended_comparative}, our proposals achieved comparable performance compared to state-of-the-art methods. To understand how close they are statistically, we performed right-tailed pairwise T-tests for results of CLAM, transformer-1, transformer-2, $\overline{H}$-w and $\overline{H}$-k128 when using SWAV-ResNet50 features. We report the $p$ values in \cref{f:p-val-1} and \cref{f:p-val-2}, \textit{rounded to 3 digits}. The $p$ values were adjusted using Benjamini/Hochberg method to account for multiple hypothesis testings.

For Normal vs Tumor in lung tissue, when using CPTAC as discovery set and TCGA as independent testing set, all method AUROC results in CPTAC-valid are not statistically different (all $p>0.05$). However, when testing on TCGA (TCGA-test), $\overline{H}$-k128 AUROC is statistically better than that of CLAM and transformer-2 (0.984±0.003 vs 0.970±0.003 and 0.975±0.002) with the $p$ values of 0.018 and 0.038 respectively. Similarly, when using TCGA as discovery set and CPTAC as independent testing set, only the results in CPTAC-test are different statistically. Specifically, $\overline{H}$-w performed as good as transformer-1 (0.9539±0.0051 vs 0.961±0.006 in AUROC) with a $p$ value of 0.076. However, it is statistically worse than CLAM and transformer-2 (0.954±0.005 vs 0.971±0.004 and 0.976±0.004 in AUROC) with the $p$ values of 0.005. On the other hand, $\overline{H}$-k128 is statistically better than transformer-1 (0.978±0.006 vs 0.961±0.006 in AUROC) with a $p$ value of 0.015. However, it performed statistically similar to CLAM and transformer-2 (0.978±0.006 vs 0.971±0.004 and 0.976±0.004 in AUROC) with the $p$ values respectively of 0.127 and 0.335.

For LUAD vs LUSC in lung tissue, all methods are not statistically different (all $p>0.05$) in their validation results (CPTAC-valid and TCGA-valid) when alternating TCGA and CPTAC as discovery cohort. When training on CPTAC and testing on TCGA (TCGA-test results), CLAM performed as well as transformer-1 and transformer-2 (0.840±0.003 vs 0.835±0.008 and 0.843±0.005 in AUROC) with the $p$ values of 0.307 and 0.356.

With respect to the results in \cref{t:extended_comparative} for subtyping cancers using only data from TCGA, the cross-validation results of all methods are not statistically different from each other for AUROC (all $p>0.05$). For mAP, in RCC, $\overline{H}-w$ and transformer-2 are statistically better than CLAM (0.983±0.003 and  0.981±0.005 vs 0.972±0.003) with the $p$ values of 0.046. Otherwise, the differences in performance of all methods are not statistically significant.

Overall, statistically speaking, our proposal performed as well as CLAM, transformer-1 and transformer-2.
}

\subsection{Extended ablation study}\label{s:extended_ablation}

{
Here, we describe additional ablation experiments that were conducted to investigate other components within the proposed framework.
}

\begin{table*}[ht!]
\caption{Evaluating the effects of using patch-level features which originate from different micron per pixel (mpp) by classifying \textcolor{blue}{Normal vs LUAD vs LUSC} {using solely lung tissue}. All WSIs (Normal+LUAD+LUSC) within the discovery cohort were utilized to derive a set of 16 prototypical patterns. These prototypical patterns were later utilized to compose $\overline{H}$-w (weighted sum of patch-level representation assigned to each pattern). Reported results are mean $\pm$ standard deviation of mAP taken across 5 stratified folds.}
\label{t:ablation_features_magnification}
\centering
\resizebox{1.0\linewidth}{!}{
\begin{NiceTabular}{cc|cc|cc|cc|cc}[]
\RowStyle{\bfseries}
& & \Block{1-4}{SWAV-ResNet50} & & & & \Block{1-4}{SUPERVISE-ResNet50} & & & \\
\toprule
\RowStyle{\bfseries}
Method & MPP & CPTAC-valid & TCGA-test & TCGA-valid & CPTAC-test & CPTAC-valid & TCGA-test & TCGA-valid & CPTAC-test \\
\midrule
 & 0.25 & 0.972±0.009 & 0.780±0.010 & 0.942±0.019 & 0.874±0.013 & {\color[HTML]{0000FF} 0.963±0.008} & {\color[HTML]{0000FF} 0.765±0.010} & {\color[HTML]{0000FF} 0.922±0.012} & 0.868±0.012 \\
\multirow{-2}{*}{transformer-1} & 0.50 & {\color[HTML]{0000FF} 0.975±0.006} & {\color[HTML]{0000FF} 0.817±0.005} & {\color[HTML]{0000FF} 0.942±0.020} & {\color[HTML]{0000FF} 0.901±0.004} & 0.961±0.004 & 0.751±0.009 & 0.909±0.014 & {\color[HTML]{0000FF} 0.871±0.009} \\
\cdashlinelr{1-11}
 & 0.25 & 0.976±0.005 & 0.725±0.013 & 0.927±0.012 & 0.723±0.023 & {\color[HTML]{0000FF} 0.965±0.009} & 0.759±0.009 & {\color[HTML]{0000FF} 0.915±0.016} & 0.791±0.017 \\
\multirow{-2}{*}{$\overline{H}$-w} & 0.50 & {\color[HTML]{0000FF} 0.977±0.002} & {\color[HTML]{0000FF} 0.809±0.008} & {\color[HTML]{0000FF} 0.942±0.015} & {\color[HTML]{0000FF} 0.858±0.006} & 0.964±0.007 & {\color[HTML]{0000FF} 0.762±0.007} & 0.906±0.008 & {\color[HTML]{0000FF} 0.817±0.009} \\
\bottomrule
\end{NiceTabular}
}
\end{table*}

\begin{figure*}[!ht]
\centering
\includegraphics[width=1.0\linewidth]{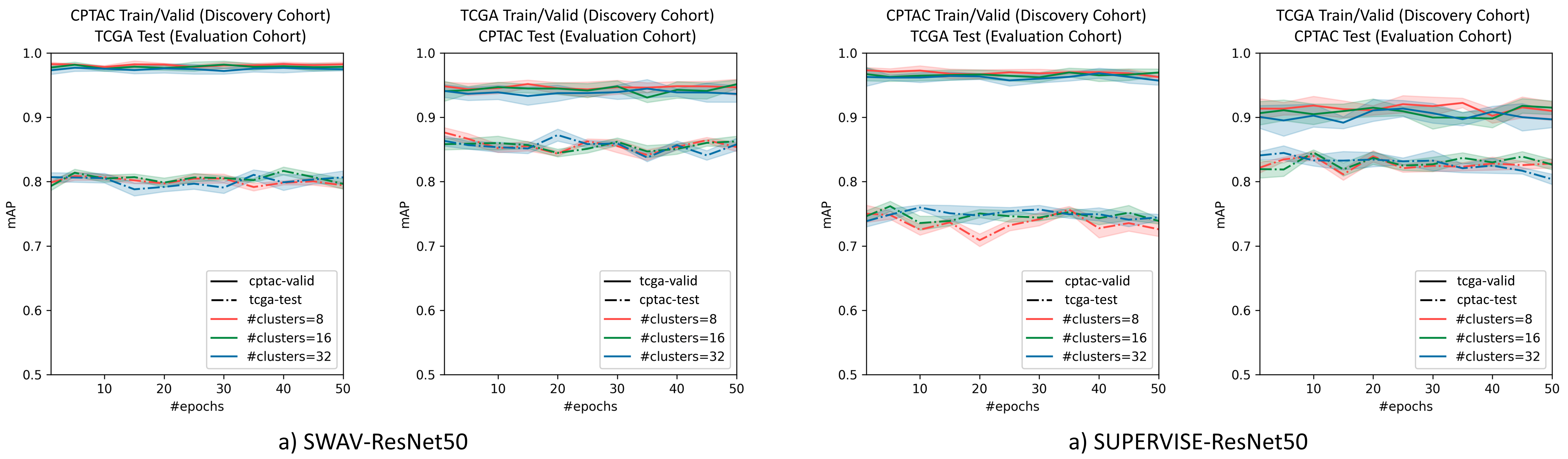}
\caption{The effect of varying number of epochs when clustering for prototypical patterns. The evaluation task is the classification of \textcolor{blue}{Normal vs LUAD vs LUSC}. All available WSIs within the discovery cohort were utilized for the clustering process. $\overline{H}$-w were utilized to compose the WSI-level H2T representation for the comparison (weighted summing patch features assigned to a pattern). Reported results are in mean $\pm$ standard deviation taken across 5 stratified folds; the shaded area denotes the error bound.}
\label{f:ablation_clustering_epochs}
\end{figure*}

\subsubsection{Clustering and the number of prototypical patterns}\label{s:ablation_clusters}

\textbf{Settings.} Clustering is the first step in our framework. Despite being conceptually simple, in practice, the usability of the resulting set patterns can only be assessed once we utilize WSI-level H2T representations derived from it for downstream tasks, such as classification for instance. Consequentially, because the nature of the downstream tasks is unknown at this derivation time, it is desirable that the clustering process could easily generate good prototypical patterns on average trials. On top of that, we also ideally want to shorten the time for obtaining the patterns as much as possible.

Within the clustering process, the number of clusters is an important parameter, especially for our H2T framework. Specifically, our WSI-level representation is a form of projection against the sets of prototypical patterns, varying the number of prototypical patterns therefore also affects the resulting WSI-level representation.

In addition to that, a good WSI-level representation should remain agnostic as much as possible with respect to their parameters. Specifically, the relative performance when using 8 patterns or 32 patterns should not be significantly better or worse compared to each other when switching from SWAV-ResNet50 to SUPERVISE-ResNet50.

With these criteria in mind, we investigate how the classification results of Normal vs LUAD vs LUSC vary under different number of clustering epochs (or iterations), number of prototypical patterns and the origins of patch-level features. By extension, these results also indicate the goodness of the prototypical patterns obtained from the k-mean clustering.

For this experiment, we assess the clustering process for obtaining 8, 16 and 32 patterns when using patch-level features of all WSIs within the discovery cohort. These patch-level features were extracted either from using SWAV-ResNet50 or SUPERVISE-ResNet50. For this experiment, we utilized $\overline{H}$-w as the WSI-level representation for the downstream classification task.

\textbf{Results.} The classification results when using either CPTAC or TCGA as the discovery cohort are provided in \cref{f:ablation_clustering_epochs}. Regardless of the patch-level feature origin, within the discovery cohort (the solid line), we observe that varying the number of clustering epochs barely affects the classification results in general. However, when using SUPERVISE-ResNet50 patch-level features for clustering 8 prototypical patterns (`\#clusters=8'), the resulting WSI-level H2T representations exhibit notable instability on the TCGA-test results (the red dashed and dotted line) when varying the number of clustering epochs.

On the other hand, by and large, the clustering processes for 16 and 32 prototypical patterns (`\#clusters=16' and `\#clusters=32' respectively) result in the WSI-level representations with a relatively similar level of performance. It is worth noting that, unless mentioned otherwise, we used 16 patterns as default experimentation setup and we used 25 as the number of clustering epochs.

Apart from that, aligning with our assumption about the agnostic level of the WSI-level representation with respect to their origins, the performance when using 8, 16 or 32 patterns maintains their relative ordering when using either SWAV-ResNet50 or SUPERVISE-ResNet50 features.

On another note, we also identify that our framework relies heavily on the origins of patch-level features. This is evident by the noticeable drop in performance on the TCGA-test (when CPTAC is the discovery cohort) and TCGA-valid (when TCGA is the discovery cohort) when moving from SWAV-ResNet50 and SUPERVISE-ResNet50. Given that our framework offers no step to enrich the patch-level feature, this outcome is expected.

Lastly, continuing the trends observed in all of the results so far, models trained using CPTAC as discovery cohort is less generalizable compared to those trained on TCGA, we speculate that the differences in the WSI extraction protocols (flash frozen vs FFPE) and/or the number of WSIs are the main reason.

\subsubsection{Effective magnification level}
\label{s:patch_size}

\textbf{Settings.} The discriminative power of the patch-level features (or representations) is of utmost importance not only for our proposed methods but also for the baseline approaches. We assume that there are two major factors that affect the patch-level representation power:

\begin{enumerate}[label=\alph*.]
\item The magnification and the shape of the image patches.
\item The CNN models that conduct the feature extraction process.
\end{enumerate}

For this experiment, because we only use patches of shape $512 \times 512$ with $256 \times 256$ degree of overlapping throughout this paper, we therefore focus on comparing features obtained from $0.50$ and $0.25$ micron per pixel (mpp) magnification instead. We again use 16 prototypical patterns and $\overline{H}$-w as the method for generating WSI-level representations for the comparison.

\textbf{Results.} The results are provided in \cref{t:ablation_features_magnification}. For the task at hand, using patch-level features coming from higher resolution has adverse effects on the generalization of all methods in general on unseen cohort.

Our proposed representation in particular performed significantly worse at $mpp=0.25$ on the evaluation set when using SWAV-ResNet50 features (a reduction of 0.08 and 0.13 in mAP respectively for TCGA-test and CPTAC-test). By tracing back the formulation of $\overline{H}$ in \cref{eq:H_formula}, it is apparent that $\overline{H}$-w or $\overline{H}$ in general, does not take much into account about the surrounding formation of the patches. Consequently, because $mpp=0.25$ is more fine-grained compared to $mpp=0.50$, it is possible that $512 \times 512$ patches at $mpp=0.50$ contain just enough contextual information about the tissue component formations whereas the former does not.

Meanwhile, although transformer-1 is also affected when using patch-level features originating from higher resolution ($mpp=0.50$), perhaps due to the encoded positional information and its trainable nature, the reduction in performance is less severe than our representation. At worst, its performance got reduced from 0.8168 to 0.7798 in mAP when using CPTAC as discovery cohort and features from SWAV-ResNet50.

Interestingly, for both approaches, while SUPERVISE-ResNet50 features performed worse than SWAV-ResNet50, they offer more stability when switching between $mpp=0.25$ and $mpp=0.50$ (with only a difference less than 0.02 in mAP on average).

\subsection{Representing WSIs using only normal tissue}

\begin{table*}[hb!]
\caption{Comparison study on classifying \textcolor{blue}{Normal vs Tumor} WSIs {using solely lung tissue}. The proposed H2T representations ($\overline{H}$ and  $\overline{C}$) were derived based on 16 prototypical patterns. These patterns in turn were obtained by using SWAV-ResNet50 patch-level features extracted from \textcolor{blue}{only Normal WSIs} within each discovery cohort. $\overline{H}$-w is obtained by weighted summing patch features assigned to a pattern; $\overline{H}$-k128 is obtained by averaging features from the top 128 closest patches assigned to a pattern; $\overline{C}$-one-hot is the representation obtained by training CNN on the one-hot-encoded pattern assignment map (PAM); $\widehat{H}$ is the histogram of the patterns within PAM; $\widehat{C}$ is the co-localization matrix of patterns within PAM. Reported results are mean $\pm$ standard deviation of AUROC taken across 5 stratified folds.}
\label{t:comparative_1_sup}
\centering
\resizebox{0.8\textwidth}{!}{
\begin{NiceTabular}{cc|cc|cc}[]
\toprule
\RowStyle{\bfseries}
Features & Method & CPTAC-valid & TCGA-test & TCGA-valid & CPTAC-test \\
\midrule
\Block{7-1}{SWAV-ResNet50} & $\widehat{H}$ & 0.924±0.014 & 0.926±0.001 & 0.892±0.008 & 0.852±0.014 \\
& $\widehat{C}$ & 0.959±0.005 & 0.893±0.006 & 0.938±0.010 & 0.861±0.011 \\
& $\widehat{H}$+$\widehat{C}$ & 0.958±0.006 & 0.889±0.005 & 0.938±0.008 & 0.868±0.009 \\
& $\overline{H}$-w & {\color[HTML]{0000FF} 0.996±0.002} & 0.971±0.002 & 0.995±0.003 & 0.934±0.007 \\
& $\overline{H}$-k128 & 0.994±0.002 & {\color[HTML]{0000FF} 0.976±0.004} & 0.996±0.002 & 0.939±0.005 \\
& $\overline{C}$-one-hot & 0.953±0.010 & 0.905±0.005 & 0.945±0.016 & 0.890±0.012 \\
& $\overline{H}$-w+$\overline{C}$-onehot & 0.995±0.002 & 0.966±0.006 & {\color[HTML]{0000FF} 0.998±0.001} & {\color[HTML]{0000FF} 0.950±0.005} \\ 
\bottomrule
\end{NiceTabular}
}
\end{table*}

\begin{table*}[hb!]
\caption{Comparison study on classifying \textcolor{blue}{LUAD vs LUSC} WSIs. The proposed H2T representations ($\overline{H}$ and  $\overline{C}$) were derived based on 16 prototypical patterns. These patterns in turn were obtained by using SWAV-ResNet50 patch-level features extracted from \textcolor{blue}{only Normal WSIs} within each discovery cohort. $\overline{H}$-w is obtained by weighted summing patch features assigned to a pattern; $\overline{H}$-k128 is obtained by averaging features from the top 128 closest patches assigned to a pattern; $\overline{C}$-one-hot is the representation obtained by training CNN on the one-hot-encoded pattern assignment map (PAM); $\widehat{H}$ is the histogram of the patterns within PAM; $\widehat{C}$ is the co-localization matrix of patterns within PAM. Reported results are mean $\pm$ standard deviation of AUROC taken across 5 stratified folds.}
\label{t:comparative_2_sup}
\centering
\resizebox{0.8\textwidth}{!}{
\begin{NiceTabular}{cc|cc|cc}[]
\toprule
\RowStyle{\bfseries}
Features & Method & CPTAC-valid & TCGA-test & TCGA-valid & CPTAC-test \\
\midrule
\Block{7-1}{SWAV-ResNet50} & $\widehat{H}$ & 0.768±0.022 & 0.592±0.002 & 0.607±0.018 & 0.656±0.004 \\
& $\widehat{C}$ & 0.855±0.018 & 0.610±0.003 & 0.678±0.025 & 0.690±0.005 \\
& $\widehat{H}$+$\widehat{C}$ & 0.852±0.007 & 0.628±0.003 & 0.670±0.017 & 0.675±0.006 \\
& $\overline{H}$-w & 0.975±0.005 & {\color[HTML]{0000FF} 0.782±0.004} & 0.897±0.022 & 0.851±0.011 \\
& $\overline{H}$-k128 & {\color[HTML]{0000FF} 0.973±0.008} & 0.744±0.006 & 0.897±0.018 & {\color[HTML]{0000FF} 0.896±0.011} \\
& $\overline{C}$-one-hot & 0.835±0.027 & 0.618±0.012 & 0.691±0.038 & 0.734±0.008 \\
& $\overline{H}$-w+$\overline{C}$-onehot & 0.973±0.008 & 0.771±0.007 & {\color[HTML]{0000FF} 0.901±0.018} & 0.862±0.005 \\
\bottomrule
\end{NiceTabular}
}
\end{table*}

Throughout the paper so far, we have focused mostly on using the prototypical patterns that were obtained based on an entire cohort. However, as our WSI-level H2T representation is in a way a projection of a WSI against the established knowledge (the prototypical patterns that we extracted), we are therefore also interested in how the WSI-level H2T representation of Tumor WSIs can be described in term of prototypical patterns obtained solely from Normal tissue. Understanding this aspect can potentially allow us to utilize the prototypical patterns to better stratify cancer grades \cite{vqdang2020unsupervised}.

\subsubsection{Visualization}
\begin{figure*}[!t]
\centering
\includegraphics[width=0.90\textwidth]{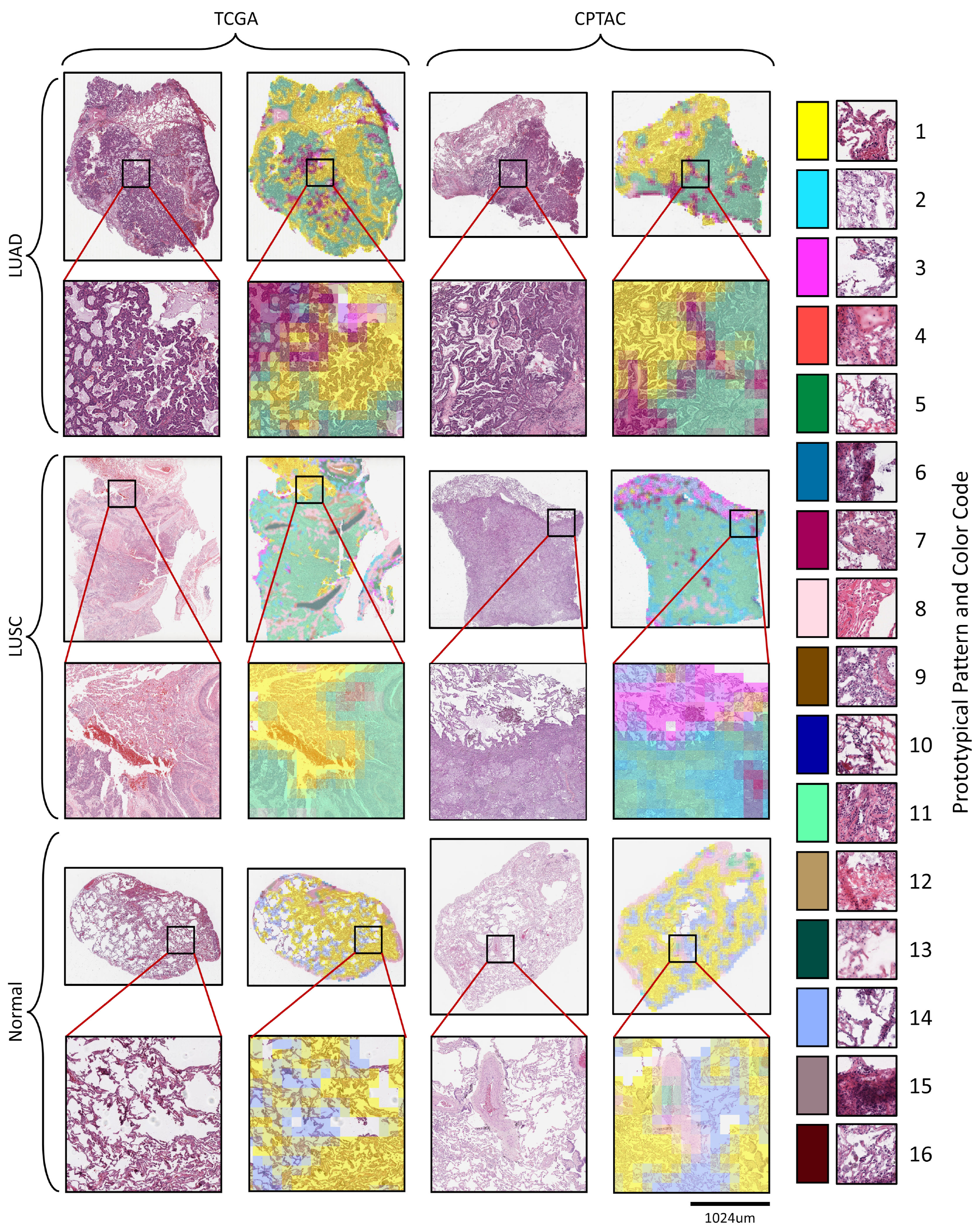}
\caption{Visual comparison of the pattern assignment maps (PAMs) between the reference cohort (Normal WSIs in TCGA) and other cohorts (Tumor WSIs in TCGA and all WSIs in CPTAC) {using solely lung tissue}. Here, PAMs were constructed using 16 prototypical patterns which were derived from TCGA cohort, using SWAV-ResNet50 patch-level features and \textcolor{blue}{only Normal WSIs} as reference tissue. Overlapping regions have their colors averaged for illustration. Locations whose colors do not align with established color code indicate the transition between assigned patterns. Note that patterns having the same color but were derived from different clustering may not be semantically similar. The colors in \cref{f:pam-pathologists} and \cref{f:pam1} denote different assignments from this figure. The color assignment is for assessing the consistency within this figure only.
}
\label{f:pam1}
\end{figure*}

Similar to what we have done, we start by examining the possible meaning of the pattern assignment maps (PAMs) when we project the WSI against a set of prototypical patterns obtained solely from Normal WSIs. We present the results in \cref{f:pam1}. For this figure, 16 prototypical patterns were extracted from Normal WSIs within the TCGA cohort using patch-level representation from SWAV-ResNet50. We observe that, in comparison to the Normal WSIs, Tumor WSIs like LUAD and LUSC are largely dominated by some specific types of prototypical pattern. Other than that, the patches assigned to the same pattern visually carry similar details. 

\subsubsection{Comparative study}

To actually assess the usability of such representation, we again perform classification of Normal vs Tumor and LUAD vs LUSC. These set of experiments follow the same setup as in \cref{h:comparative_study}. We present the results in \cref{t:comparative_1_sup} and \cref{t:comparative_2_sup}.

For Normal vs Tumor, we observe that WSI-level representation based solely on Normal WSIs remain predictive. The AUROC of $\overline{H}$-w, $\overline{H}$-k128 and $\overline{H}$-w+$\overline{C}$-one-hot in TCGA-valid are still higher than that of DSMIL-LC. In addition to that, their results in TCGA-test are not much worse compared to the performance of Transformer models in \cref{t:comparative_Normal-Tumor}.

As for LUAD vs LUSC, while the $\overline{H}$-k128 and $\overline{H}$-w+$\overline{C}$-one-hot were able to achieve AUROC more than 0.90, their performances on the evaluation set (TCGA-test and CPTAC-test) highlight their lack of generalization. As a result, it is necessary to include the samples from unknown tissue types (LUAD and LUSC) when deriving the prototypical patterns. In order to find out which WSIs containing the unknown tissue types, a sophisticated out of distribution detector may be necessary. However, as an alternative, we can also simply perform clustering on all available WSIs like what we have done so far.

Finally, we also provide a comparative study against Transformer models and CLAM. In addition to these baselines, we also compare against our own set of WSI-level representation that were obtained based on All WSIs within the discovery cohort. The results are provide in \cref{t:comparative_3_sup}.

\subsubsection{Ablation study}

We assess the effects of using different pooling strategies and present the results in \cref{t:ablation_pooling_sup}. We observe that even when using prototypical patterns based solely on Normal WSIs within the discovery cohort, the relative ordering of different weights from \cref{eq:H_formula} remains the same compared to what we have observed previously in \cref{t:ablation_pooling}. Interestingly, despite being less generalized compared to when patterns based on all WSIs, many of our representations maintain high performance in cross-validation.

\begin{table*}[!ht]
\caption{Comparison study on classifying \textcolor{blue}{Normal vs LUAD vs LUSC} WSIs {using solely lung tissue}. The proposed H2T representations ($\overline{H}$) were derived based on 16 prototypical patterns. These patterns in turn were obtained by using SWAV-ResNet50 patch-level features extracted from either only Normal WSIs or all WSIs within each discovery cohort (denoted via `Tissue Source` column). $\overline{H}$-w is obtained by weighted summing patch features assigned to a pattern; $\overline{H}$-k128 is obtained by averaging features from the top 128 closest patches assigned to a pattern. Reported results are mean $\pm$ standard deviation of AUROC taken across 5 stratified folds. For methods having zero standard deviation, the reported means are result of 1 single run.}
\label{t:comparative_3_sup}
\centering
\resizebox{1.0\textwidth}{!}{
\begin{NiceTabular}{cc|cccc|cccc}[]
\RowStyle{\bfseries}
& & \Block{1-4}{SWAV-ResNet50} & & & & \Block{1-4}{SUPERVISE-ResNet50} & & & \\
\midrule
\RowStyle{\bfseries}
Tissue Source & Method & CPTAC-valid & TCGA-test & TCGA-valid & CPTAC-test & CPTAC-valid & TCGA-test & TCGA-valid & CPTAC-test \\
\midrule
- & CLAM & 0.979±0.005 & 0.822±0.004 & 0.954±0.006 & 0.901±0.006 & {\color[HTML]{0000FF} 0.977±0.004} & {\color[HTML]{0000FF} 0.789±0.013} & 0.933±0.012 & {\color[HTML]{0000FF} 0.893±0.0032} \\
- & transformer-1 & 0.975±0.006 & 0.817±0.005 & 0.942±0.020 & {\color[HTML]{0000FF} 0.901±0.004} & 0.961±0.004 & 0.751±0.009 & 0.909±0.014 & 0.871±0.0094 \\
- & transformer-2 & {\color[HTML]{0000FF} 0.981±0.004} & {\color[HTML]{0000FF} 0.828±0.009} & {\color[HTML]{0000FF} 0.958±0.010} & 0.899±0.009 & 0.974±0.005 & 0.786±0.010 & {\color[HTML]{0000FF} 0.946±0.007} & 0.887±0.0043 \\
\midrule
Normal & $\overline{H}$-w & {\color[HTML]{0000FF} 0.973±0.004} & {\color[HTML]{0000FF} 0.793±0.006} & {\color[HTML]{0000FF} 0.930±0.013} & 0.798±0.019 & 0.963±0.006 & 0.721±0.006 & 0.893±0.014 & {\color[HTML]{0000FF} 0.825±0.0110} \\
Normal & $\overline{H}$-k128 & 0.973±0.007 & 0.770±0.008 & 0.929±0.015 & {\color[HTML]{0000FF} 0.825±0.010} & {\color[HTML]{0000FF} 0.969±0.006} & {\color[HTML]{0000FF} 0.722±0.006} & {\color[HTML]{0000FF} 0.906±0.026} & 0.806±0.0136 \\
\midrule
Normal+LUAD+LUSC & $\overline{H}$-w & 0.977±0.002 & 0.809±0.008 & 0.942±0.015 & 0.858±0.006 & 0.964±0.007 & 0.762±0.007 & 0.906±0.008 & 0.817±0.0094 \\
Normal+LUAD+LUSC & $\overline{H}$-k128 & {\color[HTML]{0000FF} 0.984±0.004} & {\color[HTML]{0000FF} 0.825±0.005} & {\color[HTML]{0000FF} 0.957±0.014} & {\color[HTML]{0000FF} 0.890±0.007} & {\color[HTML]{0000FF} 0.979±0.005} & {\color[HTML]{0000FF} 0.775±0.001} & {\color[HTML]{0000FF} 0.933±0.019} & {\color[HTML]{0000FF} 0.855±0.0120} \\
\bottomrule
\end{NiceTabular}
}
\end{table*}

\begin{table*}[t!]
\caption{Ablation study on different pooling strategies from \cref{eq:H_formula} and the impacts from using different patch-level representations on the proposed constructions. Here, the task is classifying \textcolor{blue}{Normal vs LUAD vs LUSC} {using solely lung tissue}. Only \textbf{Normal} WSIs within the discovery cohort were utilized to derive 16 prototypical patterns. Reported results are mean $\pm$ standard deviation of mAP taken across 5 stratified folds.}
\label{t:ablation_pooling_sup}
\centering
\resizebox{\linewidth}{!}{
\begin{NiceTabular}{c|cc|cc|cc|cc}[]

\RowStyle{\bfseries}
& \Block{1-4}{SWAV-ResNet50} & & & & \Block{1-4}{SUPERVISE-ResNet50} & & & \\
\toprule

\RowStyle{\bfseries}
Method & CPTAC-valid & TCGA-test & TCGA-valid & CPTAC-test & CPTAC-valid & TCGA-test & TCGA-valid & CPTAC-test \\
\midrule
$\overline{H}$ & 0.972±0.003 & 0.790±0.004 & 0.924±0.014 & 0.795±0.017 & {\color[HTML]{0000FF} 0.964±0.007} & 0.716±0.004 & {\color[HTML]{0000FF} 0.896±0.010} & 0.824±0.010\\
$\overline{H}$-w & {\color[HTML]{0000FF} 0.973±0.004} & {\color[HTML]{0000FF} 0.793±0.006} & {\color[HTML]{0000FF} 0.930±0.013} & {\color[HTML]{0000FF} 0.798±0.019} & 0.963±0.006 & {\color[HTML]{0000FF} 0.721±0.006} & 0.893±0.014 & {\color[HTML]{0000FF} 0.825±0.011} \\
\cdashlinelr{1-10}
$\overline{H}$-t0.2 & 0.972±0.003 & {\color[HTML]{0000FF} 0.790±0.004} & {\color[HTML]{0000FF} 0.924±0.014} & {\color[HTML]{0000FF} 0.795±0.017} & {\color[HTML]{0000FF} 0.964±0.007} & 0.716±0.004 & {\color[HTML]{0000FF} 0.897±0.010} & 0.824±0.010\\
$\overline{H}$-t0.3 & {\color[HTML]{0000FF} 0.974±0.005} & 0.787±0.006 & 0.924±0.014 & 0.793±0.015 & 0.963±0.010 & {\color[HTML]{0000FF} 0.724±0.010} & 0.894±0.014 & {\color[HTML]{0000FF} 0.825±0.011} \\
$\overline{H}$-t0.4 & 0.972±0.006 & 0.786±0.006 & 0.922±0.010 & 0.785±0.019 & 0.954±0.006 & 0.710±0.007 & 0.870±0.013 & 0.786±0.011\\
$\overline{H}$-t0.5 & 0.916±0.006 & 0.724±0.009 & 0.855±0.011 & 0.707±0.012 & 0.825±0.014 & 0.641±0.008 & 0.761±0.010 & 0.645±0.008\\
$\overline{H}$-t0.6 & 0.728±0.010 & 0.511±0.006 & 0.642±0.013 & 0.637±0.007 & 0.651±0.021 & 0.450±0.006 & 0.519±0.015 & 0.547±0.011\\
$\overline{H}$-t0.7 & 0.472±0.019 & 0.387±0.004 & 0.464±0.020 & 0.511±0.006 & 0.472±0.020 & 0.380±0.003 & 0.392±0.012 & 0.463±0.006\\
\cdashlinelr{1-10}
$\overline{H}$-k8 & 0.953±0.010 & 0.704±0.007 & 0.900±0.014 & 0.750±0.010 & 0.936±0.009 & 0.673±0.007 & 0.855±0.025 & 0.729±0.026\\
$\overline{H}$-k16 & 0.959±0.008 & 0.727±0.006 & 0.905±0.015 & 0.779±0.010 & 0.951±0.011 & 0.695±0.004 & 0.882±0.028 & 0.766±0.020\\
$\overline{H}$-k32 & 0.964±0.007 & 0.744±0.007 & 0.916±0.021 & 0.792±0.010 & 0.958±0.007 & 0.708±0.002 & 0.890±0.025 & 0.787±0.014\\
$\overline{H}$-k64 & 0.966±0.006 & 0.757±0.007 & 0.922±0.015 & 0.810±0.010 & 0.965±0.007 & 0.718±0.004 & 0.898±0.028 & 0.801±0.017\\
$\overline{H}$-k128 & {\color[HTML]{0000FF} 0.973±0.007} & {\color[HTML]{0000FF} 0.770±0.008} & {\color[HTML]{0000FF} 0.929±0.015} & {\color[HTML]{0000FF} 0.825±0.010} & {\color[HTML]{0000FF} 0.969±0.006} & {\color[HTML]{0000FF} 0.722±0.006} & {\color[HTML]{0000FF} 0.906±0.026} & {\color[HTML]{0000FF} 0.806±0.014} \\
\cdashlinelr{1-10}
$\overline{H}$-fk8 & 0.905±0.017 & 0.688±0.006 & 0.839±0.017 & 0.664±0.018 & 0.872±0.017 & 0.613±0.014 & 0.835±0.007 & 0.686±0.014\\
$\overline{H}$-fk16 & 0.914±0.024 & 0.704±0.005 & 0.859±0.014 & 0.691±0.014 & 0.884±0.012 & 0.622±0.014 & 0.850±0.008 & 0.704±0.011\\
$\overline{H}$-fk32 & 0.927±0.015 & 0.727±0.005 & 0.873±0.008 & 0.716±0.011 & 0.896±0.017 & 0.633±0.015 & 0.863±0.006 & 0.715±0.010\\
$\overline{H}$-fk64 & 0.937±0.012 & 0.742±0.006 & 0.886±0.007 & 0.731±0.015 & 0.903±0.019 & 0.652±0.013 & 0.876±0.007 & 0.728±0.009\\
$\overline{H}$-fk128 & {\color[HTML]{0000FF} 0.946±0.008} & {\color[HTML]{0000FF} 0.748±0.007} & {\color[HTML]{0000FF} 0.899±0.007} & {\color[HTML]{0000FF} 0.749±0.014} & {\color[HTML]{0000FF} 0.922±0.015} & {\color[HTML]{0000FF} 0.676±0.011} & {\color[HTML]{0000FF} 0.887±0.014} & {\color[HTML]{0000FF} 0.742±0.009} \\
\bottomrule
\end{NiceTabular}
}
\end{table*}

\begin{table*}[t!]
\caption{Ablation study on the proposed H2T representations based on the co-localization of prototypical patterns by classifying \textcolor{blue}{Normal vs LUAD vs LUSC} {using solely lung tissue}. The prototypical patterns were derived using patch-level features from \textcolor{blue}{SWAV-ResNet50}. We also evaluate the effects of using different sources of WSIs on such derivation: using only Normal WSIs or all WSIs (Normal+LUAD+LUSC) within the discovery cohort. $\overline{C}$-raw is the representation obtained by training CNN on the pattern assignment map (PAM); $\overline{C}$-one-hot is the representation obtained by training CNN on the one-hot-encoded PAM; $\widehat{C}$ is the co-localization matrix of patterns within PAM. Reported results are in mean $\pm$ standard deviation taken across 5 stratified folds.}
\label{t:ablation_colocalization_Normal-LUAD-LUSC}
\centering
\resizebox{\textwidth}{!}{
\begin{NiceTabular}{cc|cccc|cccc}[]
\toprule
\RowStyle{\bfseries}
& & \Block{1-4}{Tissue Source - Normal} & & & & \Block{1-4}{Tissue Source - Normal+LUAD+LUSC} & & & \\
\midrule
\RowStyle{\bfseries}
{\#}clusters & Method & CPTAC-valid & TCGA-test & TCGA-valid & CPTAC-test & CPTAC-valid & TCGA-test & TCGA-valid & CPTAC-test \\
\midrule
8 & $\widehat{C}$ & {\color[HTML]{0000FF} 0.791±0.023} & 0.538±0.003 & 0.615±0.011 & 0.507±0.006 & {\color[HTML]{0000FF} 0.815±0.015} & 0.570±0.003 & 0.690±0.021 & 0.673±0.010 \\
8 & $\overline{C}$-raw & 0.779±0.016 & 0.508±0.007 & 0.591±0.009 & 0.533±0.008 & 0.757±0.012 & 0.531±0.014 & 0.626±0.023 & 0.649±0.022 \\
8 & $\overline{C}$-one-hot & 0.772±0.025 & {\color[HTML]{0000FF} 0.551±0.019} & {\color[HTML]{0000FF} 0.648±0.017} & {\color[HTML]{0000FF} 0.641±0.022} & 0.770±0.038 & {\color[HTML]{0000FF} 0.604±0.010} & {\color[HTML]{0000FF} 0.691±0.012} & {\color[HTML]{0000FF} 0.673±0.008} \\
\cdashlinelr{1-10}
16 & $\widehat{C}$ & 0.833±0.020 & 0.546±0.006 & {\color[HTML]{0000FF} 0.679±0.023} & 0.575±0.008 & {\color[HTML]{0000FF} 0.857±0.019} & {\color[HTML]{0000FF} 0.617±0.003} & {\color[HTML]{0000FF} 0.741±0.026} & {\color[HTML]{0000FF} 0.691±0.004} \\
16 & $\overline{C}$-raw & 0.786±0.016 & 0.483±0.009 & 0.537±0.017 & 0.540±0.028 & 0.779±0.032 & 0.487±0.005 & 0.594±0.025 & 0.592±0.027 \\
16 & $\overline{C}$-one-hot & 0.823±0.040 & {\color[HTML]{0000FF} 0.569±0.012} & 0.669±0.032 & {\color[HTML]{0000FF} 0.656±0.020} & 0.784±0.037 & 0.597±0.013 & 0.700±0.039 & 0.672±0.014 \\
\cdashlinelr{1-10}
32 & $\widehat{C}$ & {\color[HTML]{0000FF} 0.866±0.005} & 0.562±0.006 & {\color[HTML]{0000FF} 0.716±0.028} & 0.599±0.006 & {\color[HTML]{0000FF} 0.911±0.018} & {\color[HTML]{0000FF} 0.636±0.007} & {\color[HTML]{0000FF} 0.776±0.011} & {\color[HTML]{0000FF} 0.716±0.008} \\
32 & $\overline{C}$-raw & 0.725±0.040 & 0.506±0.008 & 0.577±0.027 & 0.614±0.014 & 0.803±0.050 & 0.471±0.010 & 0.548±0.039 & 0.566±0.015 \\
32 & $\overline{C}$-one-hot & 0.837±0.036 & {\color[HTML]{0000FF} 0.567±0.009} & 0.693±0.006 & {\color[HTML]{0000FF} 0.629±0.024} & 0.879±0.040 & 0.590±0.017 & 0.758±0.017 & 0.716±0.022 \\
\bottomrule
\end{NiceTabular}
}
\end{table*}

\end{document}